\documentclass[11pt,pre,onecolumn,tightenlines,longbibliography,notitlepage]{revtex4-1}
\pdfoutput=1

\usepackage{xcolor}
\usepackage{verbatim}
\usepackage{graphicx}

\newcommand\hl[1]{%
  \bgroup
  \hskip0pt\color{red!80!black}%
  #1%
  \egroup
}

\begin{document}

\title{Diffusion-driven transition between two regimes of viscous fingering}

\author{Thomas E. Videb\ae k}
\affiliation{Department of Physics and the James Franck and Enrico Fermi Institutes, The University of Chicago, Chicago IL 60637}
\author{Sidney R. Nagel} 
\affiliation{Department of Physics and the James Franck and Enrico Fermi Institutes, The University of Chicago, Chicago IL 60637}

\begin{abstract}

Viscous fingering patterns can form at the interface between two immiscible fluids confined in the gap between a pair of flat plates; whenever the fluid with lower viscosity displaces the one of higher viscosity the interface is unstable.  For miscible fluids the situation is more complicated due to the formation of interfacial structure in the thin dimension spanning the gap. Here we study the effect of the inherent diffusion between the two miscible fluids on this structure and on the viscous fingering patterns that emerge.  We discover an unexpected transition separating two distinct regimes  where the pattern morphologies and mode of onset are different.  This transition is marked by a regime of transient stability as the structure of the fingers evolves from having three-dimensional structure to being quasi-two dimensional. The presence of diffusion allows an instability to form where it was otherwise forbidden.

\end{abstract}

\maketitle

\section{Introduction}

Diffusion normally acts to make a system more uniform.  Over time, diffusion smears out and removes structures that have appeared and increases the length scale on which patterns can form.   Unless there are chemically reactive components as in a Turing instability~\cite{Turing1952,Kondo2010}, one might naively expect that diffusion would stabilize a system against instabilities.  However, for viscous fingering we find that while diffusion does make the system more stable {\em initially}, 
it surprisingly allows new instabilities to appear at a sharp transition when it acts on longer time scales.  

\begin{figure}
	\centering
	\includegraphics[width=\linewidth]{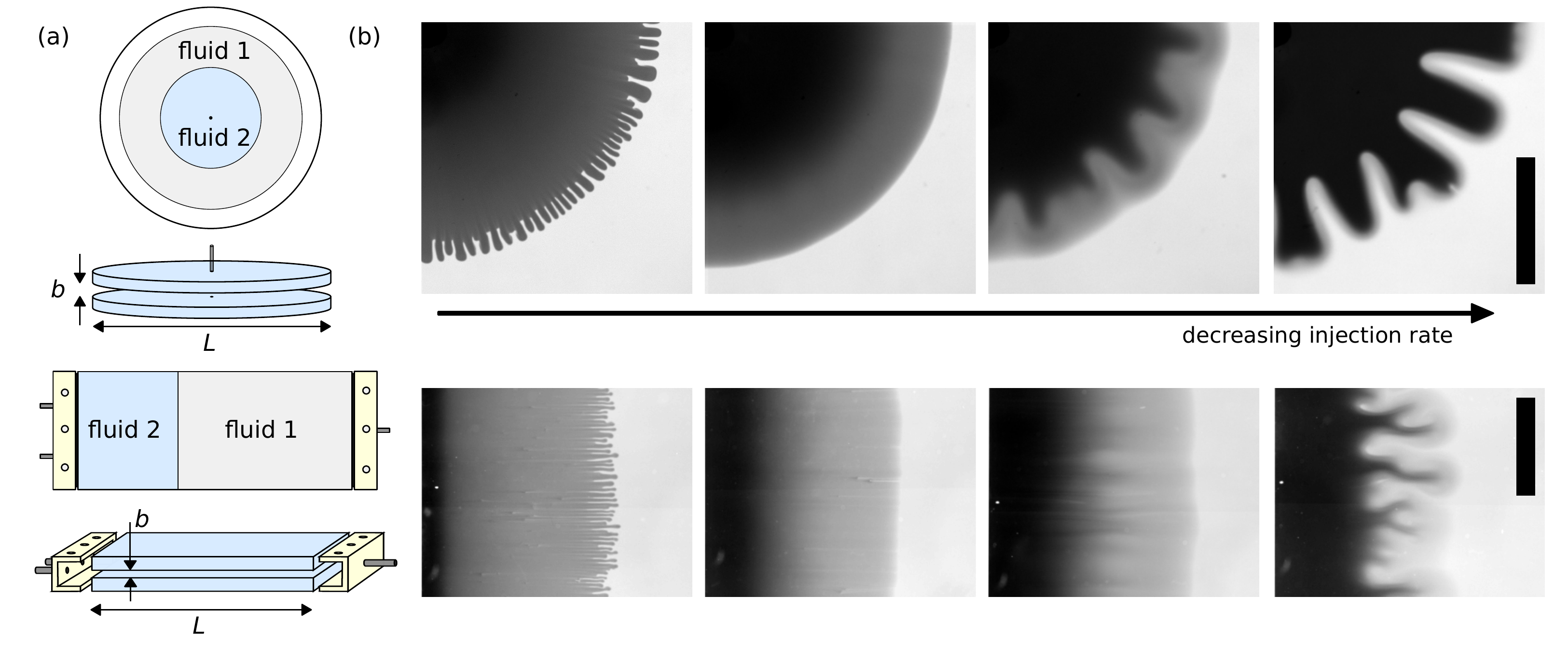}
	\caption{ (a) Schematic of both a radial and rectilinear Hele-Shaw cell showing two large flat plates, of size $L$, separated by a small gap of width $b$. (b) Fingering patterns for different injection rates. One of the fluids is dyed and the grey level indicates the local concentration of inner-fluid. Both scale bars are 2.5cm. For the radial cell (top row of images) $b=205 \mathrm{\mu m}$ and for linear cell (bottom row of images) $b=356 \mathrm{\mu m}$, both have $\eta_{in}/\eta_{out}=0.2$. (Note that the images for the linear cell had the outer fluid dyed; here we invert the colors for comparison.)}
     \label{fig:IntroTransition}
\end{figure}

The viscous fingering instability occurs when one fluid displaces another of higher viscosity in a narrow geometry, such as a Hele-Shaw cell illustrated in Fig.~\ref{fig:IntroTransition}a, where the fluid is confined in a thin gap between two large flat plates.  This is an instance of complex structure formation from benign initial conditions and has been a prototypical example of pattern formation \cite{Bensimon86,Homsy87} since the work of Saffman and Taylor in 1958 \cite{Saffman58}.  They derived a most unstable wavelength, $\lambda_c$, at which scale the interface between the fluids should be unstable. Understanding pattern formation instigated by a dynamic instability remains an ongoing challenge~\cite{Nagel2017}.

In the limit that $\lambda_c \rightarrow 0$, which can be approached by using fluids with low interfacial tensions or rapidly moving interfaces, it has been predicted that highly ramified patterns emerge and that singularities should form; the global pattern should be similar to structures seen in diffusion-limited aggregation \cite{Witten1981,Paterson84DLA,Sander85,Nittmann1985,Sander86} and the protruding fingers should form cusps at their tips \cite{Wiegmann05}. Experimental work \cite{Swinney05} was able to confirm the fractal geometry but did not observe cusps in fluid systems although they were observed in granular Hele-Shaw experiments \cite{Cheng08}. In this limit of very small $\lambda_c$, a variety of counter-intuitive phenomena have been investigated in experiments using miscible fluids with ultra-low interfacial tensions. In this case, it is no longer possible to treat the flow as purely two-dimensional in the plane parallel to the plates - rather, structures in the third dimension spanning the gap play an important role in forming the patterns. 

Experimentally no lateral (\textit{i.e.}, two-dimensional as viewed from above) fingers are observed in miscible fluids when the ratio of the inner-fluid viscosity, $\eta_{in}$, to that of the outer fluid, $\eta_{out}$, is sufficiently large but still in a regime where immiscible fluids finger readily: $0.33<\eta_{in}/\eta_{out} <1$ \cite{Lajeunesse97,Lajeunesse99,Bischofberger14}. In this stable regime of miscible fluids, the inner fluid forms a tongue with a nearly parabolic profile spanning the gap that gradually tapers down as it approaches its tip.  In contrast, when $\eta_{in}/\eta_{out} <0.33$ the profile is blunt; the inner fluid half fills the gap and only near its tip does its thickness decrease rapidly to zero.  Lajeunesse \textit{et al.} showed that the transition to lateral fingering at the cross-over viscosity ratio is marked by the tip shape changing from rounded to blunt~\cite{Lajeunesse99}. Despite additional work exploring this thickness profile~\cite{Bischofberger14,Goyal06,Oliveira11} there has been no complete explanation of why the onset of a blunt tip coincides with the lateral fingering instability. 
However, if a blunt interface is indeed a necessary condition for the lateral instability, it suggests that disrupting the tip structure might lead to stable evolution.  In this work, we explore this possibility. When we slow down the injection rate, so that diffusion has time to blur the interface substantially, the degree of fingering first decreases and then undergoes a sharp unexpected transition into a new regime.

Figure~\ref{fig:IntroTransition}b shows images taken at different injection rates for both a radial and a rectilinear Hele-Shaw experiment with the same pair of miscible fluids. All the images are taken when the outermost interface of the injected fluid has reached 5 cm from the inlet. At the left, when the injection rate is the largest, a fringe of stubby fingers is clearly evident at the outer pattern edge.  As the injection rate is decreased, the length of these fingers decreases until in the second image they completely disappear. At this point, the evolution appears to be completely stable. 

The final images of Fig.~\ref{fig:IntroTransition}b show that as the injection rate is decreased even further the fingers reemerge and grow longer with decreasing injection rate.  The finger morphology is qualitatively different in this novel, low-injection-rate regime: the fingers have a larger lateral width and appear more uniformly colored than their counterparts at higher injection rate.  As we will show, this uniform coloration corresponds to the inner invading fluid filling the gap almost entirely.  It suggests that in this regime the three-dimensional structure between the plates has been largely eradicated and the system has become quasi-two dimensional.  

A sharp transition between two distinct types of patterns provides a particularly effective condition for probing the underlying physics Ð similar to the role that thermodynamic phase transitions provide in giving a deeper perspective on the phases of matter.
The novel transition we observe occurs when fluid advection is still highly dominant over diffusion.  Fingering, albeit of different forms, occurs on both sides of the transition. The role of diffusion is not 
so dominant that all structure formation is prevented~\cite{Tan87,Yortsos87}. The demonstration that there is an intermediate regime of transient stability, where diffusion wipes out structure only in the smallest dimension 
provides insight into how different features of the miscible Hele-Shaw problem are related.

\section{Methods} 

\begin{figure}[b]
	\centering
	\includegraphics[width=3.4in]{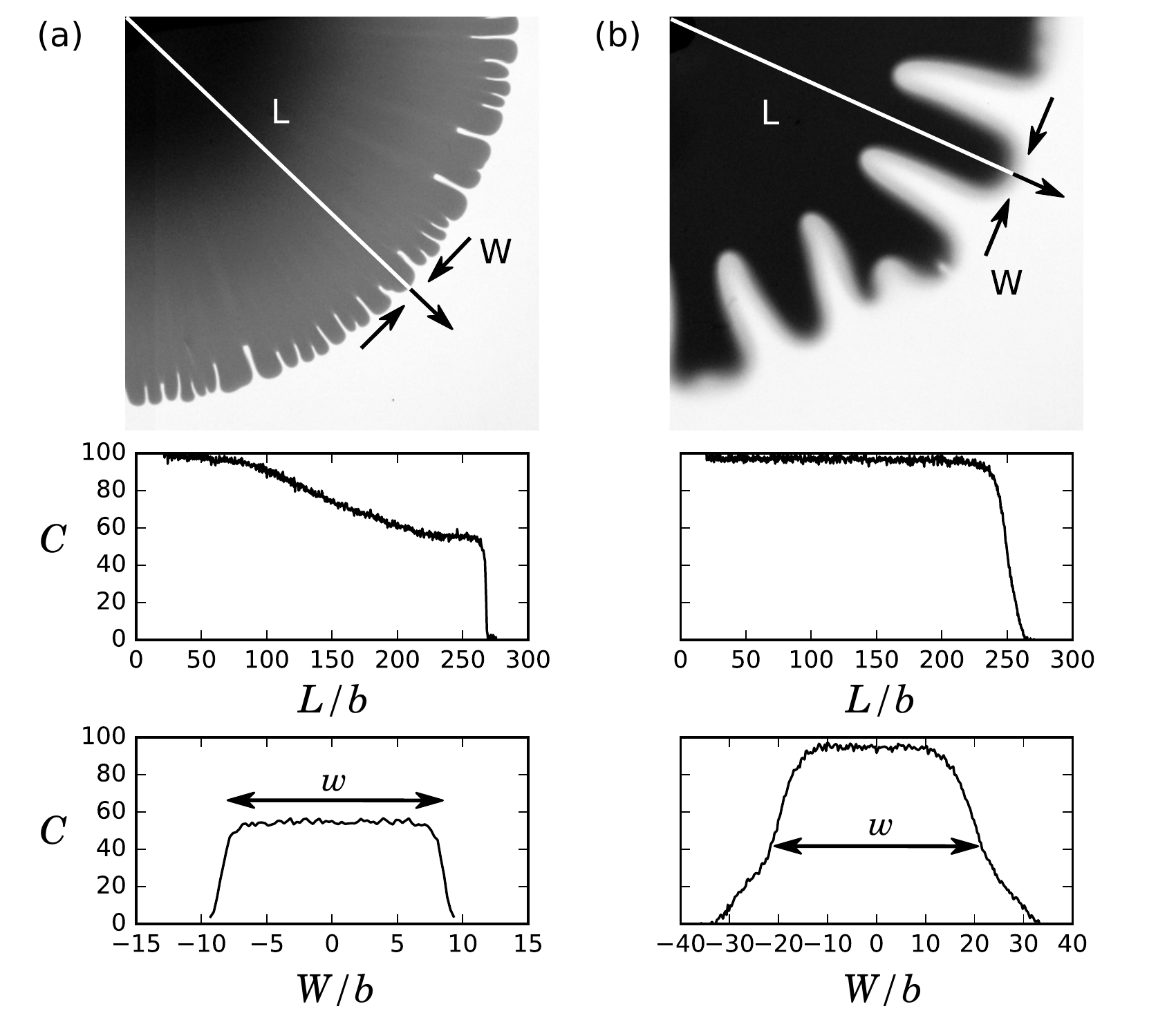}
	\caption{Profiles of gap-averaged inner fluid concentrations, $C$, from (a) high and (b) low injection-rate experiments taken along the lines L and W. Tip-splitting occurs when a fingers width, denoted by $w$, reaches a value of $2\lambda_c$.}
     \label{fig:Measurement}
\end{figure}

The radial Hele-Shaw cell consists of two large, flat glass plates of 1.9 cm thickness and 14 cm radius with a uniform gap, $b$, between them. This gap is varied by inserting spacers of varying thickness: $76 \mu$m $< b < 419 \mu$m. There are inlets at the center of both the top and bottom plates, one for injection of pairs of fluid and the other for the removal of air bubbles; fluids used are primarily water-glycerol mixtures. 
Before beginning the experiment we flow the inner fluid into the waste tube, clearing any residual outer fluid from the inlet. The injection rate is precisely controlled by a syringe pump (NE-1000 from New Era Pump Systems Inc.); the rates used vary from 0.001 mL/min to 10 mL/min. 
Fluids are dyed with brilliant blue G from Alfa Aesar; concentrations of dye in fluids are 0.4 mg/mL. Viscosities of fluids are measured using the SVM 3001 viscometer and MCR 301 rheometer from Anton Paar. To measure the gap-averaged concentration, $C$, of the inner fluid we dye our fluids and compare the intensity of the pattern to a calibrated cell of known thickness. Figure~\ref{fig:Measurement} shows inner-fluid profiles in the high and low injection-rate regimes.

One feature of a radial geometry is that the fluid velocity is inversely proportional to the distance from the inlet. To check the effect of geometry we also conducted experiments in a rectilinear cell where the velocity of the fluid interface does not depend on the distance from the inlet. 
As we will show, the characteristics of the transition between the two fingering regimes remain in this linear geometry.

The rectilinear cell is made of two glass plates that are 1.9 cm in thickness and 17.8 cm by 30.5 cm in width and length. The gap of the cell is set by inserting spacers of the desired thickness and a seal is made on the side using silicone rubber of slightly larger thickness than the final gap height; clamping the plates compresses the rubber and seals the cell. To achieve a uniform velocity profile across the cell a reservoir is placed on either end, with an o-ring seal to secure it against the glass. These reservoirs have air holes that can be sealed on their tops.  These allow complete filling of the reservoir without inducing pressure gradients within the cell. The reservoirs are 1.3 cm in height, 15.4 cm in length, and 1.3 cm in depth on the interior. There are then two injection points that are connected to syringe pumps. The purpose of this larger volume is to smooth out the pressure from the two point injections so as to ensure a uniform front in the cell.

The main challenge with the rectilinear cell is to create a clean initial profile. In an attempt to accomplish this the heavier outer fluid is filled while the cell is held at a 60$^{\circ}$ angle from horizontal. When the outer fluid reaches the end of the cell the fluid is left to settle under gravity until a flat interface forms. At this point the inner fluid reservoir is attached and the lower viscosity fluid is slowly added until the chamber is full. As the reservoir fills and the inner fluid comes into contact with the outer fluid, capillary forces draw the two fluids together. This prevents large air bubbles from forming. There are a few minutes where the two fluids are in contact and can diffuse into each other at this interface. As we will show, this initial diffusion does not effect the phenomena we report here. After the inner fluid reservoir is fully filled the cell is lowered to a horizontal position and the experiment can begin. 

An additional complication arising from this loading procedure is that at the small-gap inlet the fluids can pin or small air bubbles can form. These lead to defects that can disrupt a clean interface. Fortunately they do not have long-range lateral effects along the interface between the fluids and we observe the interface where there is a clean section.

To characterize the patterns we measure several parameters: (i) The most unstable wavelength, $\lambda_c$, is measured either at the onset of the instability or at tip-splitting events (when a single finger splits into two). At low injection rates a diffusive layer builds up along the edge of the pattern. To measure the width of fingers in this regime only the thicker region is considered as shown in Fig.~\ref{fig:Measurement}b. (ii) The length of fingers, $L_{finger} \equiv L_{out}-L_{in}$, is the difference between an outer length, $L_{out}$, and an inner length, $L_{in}$. For the radial cell $L_{out}$ is the radius of the smallest circle that encloses the entire pattern and $L_{in}$ is the radius of the largest circle that fits within the fully displaced region on the pattern's interior~\cite{Bischofberger14}. For the linear cell $L_{out}$ is the furthest distance from the inlet that any fingers have reached and $L_{in}$ is the shortest distance from the inlet that reaches the interface of the pattern. (iii) The instability onset (both onset length, $L_{onset}$, and onset time, $t_{onset}$) is measured by tracking a finger's length back in time to determine when it was first formed. This can either be the radius for the radial cell or a distance from the inlet for the linear cell. (iv)  The initial dimensionless growth rate, $\Gamma_{init} \equiv d(L_{finger}/L_{in})/d(L_{in}/b)$ is the growth rate of the fingers at the moment they form.

The width of the interface is described by a mutual diffusion coefficient that depends on the local concentration of inner and outer liquids as they mix.  In Appendix A, we describe how to extract an effective diffusion constant, $D$, that approximates the growth of asymmetric concentration profiles.

The diffusion of the dyes used to enhance the optical contrast between the fluids is typically slower than that of the fluids themselves; for our dyed water-glycerol mixtures, it is $O(10^2)$ smaller than the inter-diffusion of the water-glycerol mixtures. To demonstrate that the observed patterns are not an artifact of using dyes, we have also used schlieren optics that exploits the small, natural index-of-refraction difference between the fluids to measure the location of the interface~\cite{Bunton2016}. Using this technique as described in Appendix B, we find only a slight difference in the concentration profiles measured by the dye and by where the schlieren setup images a gradient in the index of refraction; the diffuse region in the dye extends only a bit further into the pattern showing that advection is high enough so that the dye and fluid remain well mixed.  Moreover, we find consistent results when the outer (rather than the inner) fluid is dyed.


To be sure that the transition we are seeing is not an effect of gravitational forces 
we check to see whether the glycerol-water experiments could be effected by gravity. Specifically we check if this is the case at onset, since this is where we measure different observables. 
We construct a dimensionless number, $F$, that measures the relative importance of gravity. $F$ is the ratio between the local interfacial velocity ($V$) and a settling velocity, which we take here to scale like the Stoke's velocity for an object of size $b$: ($\Delta\rho g b^2/ \eta_{out}$). This gives:
\begin{equation}
F = (\Delta\rho g b^2) / (\eta_{out}V )   
\end{equation}
Calculating the largest observed $F$ for each data set we find values bewteen $10^{-2}$ and $10^{-6}$.  See Appendix C for a table of experimental values. From this we conclude that viscous forces are dominant in the experiments we have performed and that gravity can be ignored for the phenomena observed. 

\section{Experimental Results}
 
The images in Fig.~\ref{fig:IntroTransition}b show a striking change in the fingering patterns as the injection rate is varied.  Figures~\ref{fig:Measurement}a and b show that the patterns formed in the two extremes have different three-dimensional profiles. At high injection rate, we find profiles in accord with those observed in previous studies \cite{Lajeunesse97,Lajeunesse99,Bischofberger14}: near their tips, fingers are blunt and the inner fluid fills approximately half the gap. In contrast, at low injection rates the fingers begin to fill the gap more fully out to their tips.

\begin{figure}
	\centering
	\includegraphics[width=3.9in]{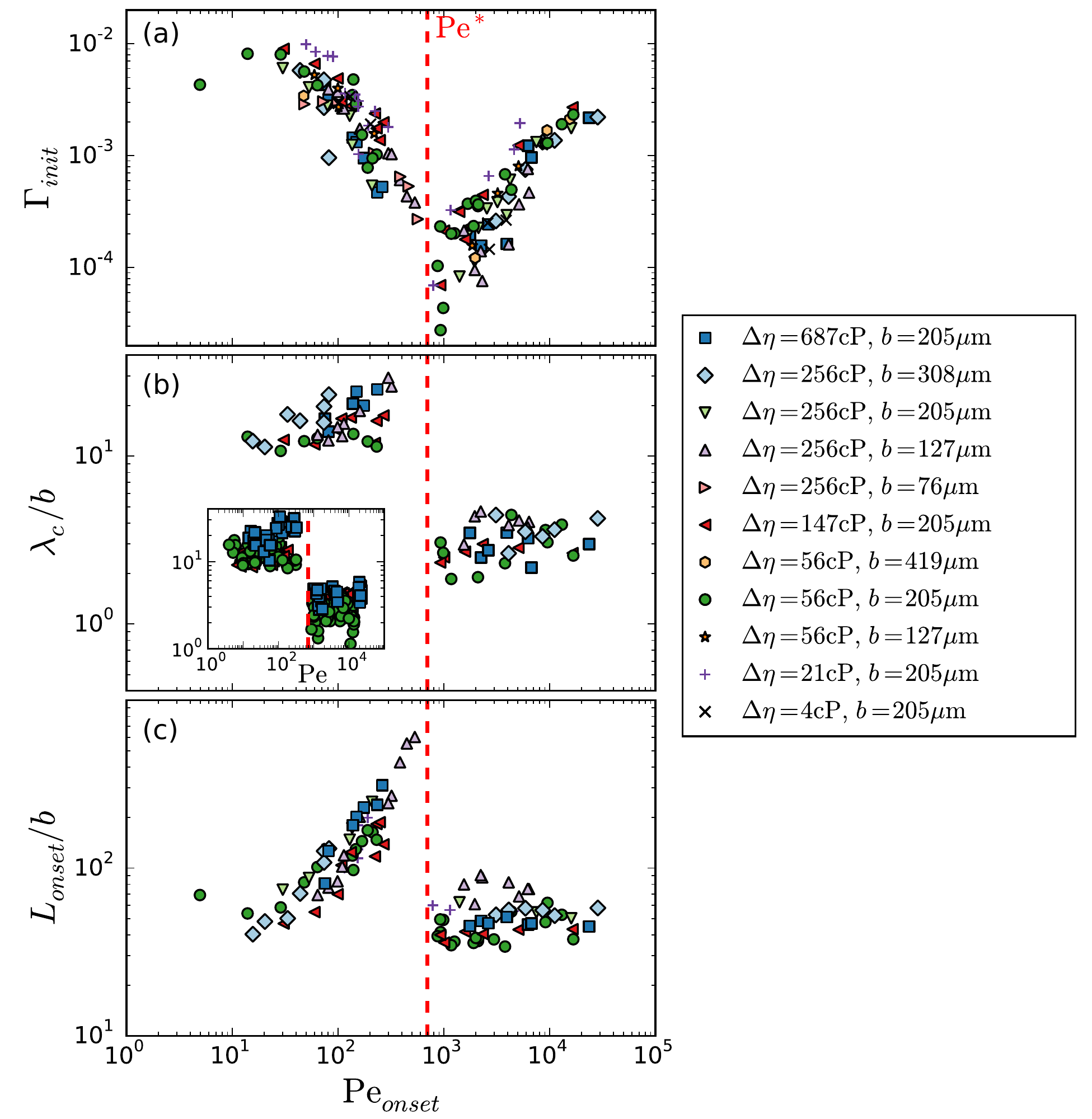}
	\caption{(a) Finger growth rate, $\Gamma_{init}$, (b) 
	normalized most unstable wavelength, $\lambda_c/b$, and (c) normalized onset length, $L_{onset}/b$, are shown versus $\mathrm{Pe}_{onset}$. Inset of (b) shows $\lambda_c/b$ measured away from onset. At a critical value, $\mathrm{Pe}^*$ (the dotted line), there is a smooth decrease in $\Gamma_{init}$, and a sharp jump in both $\lambda_c/b$ and $L_{onset}/b$.
	}
     \label{fig:CombinedTransition_RfRi}
\end{figure} 

To account for the competition between diffusion and advection in determining the shape of the interface, we introduce a dimensionless P\'eclet number defined as $\mathrm{Pe}=Vb/D$, where $V$ is the local fluid velocity at the interface, $b$ is the plate spacing, and $D$ is the effective inter-fluid diffusivity. We also define $\mathrm{Pe}_{onset}$, which is $\mathrm{Pe}$ at the onset of the fingering instability. 

\begin{figure}
	\centering
	\includegraphics[width=2.6in]{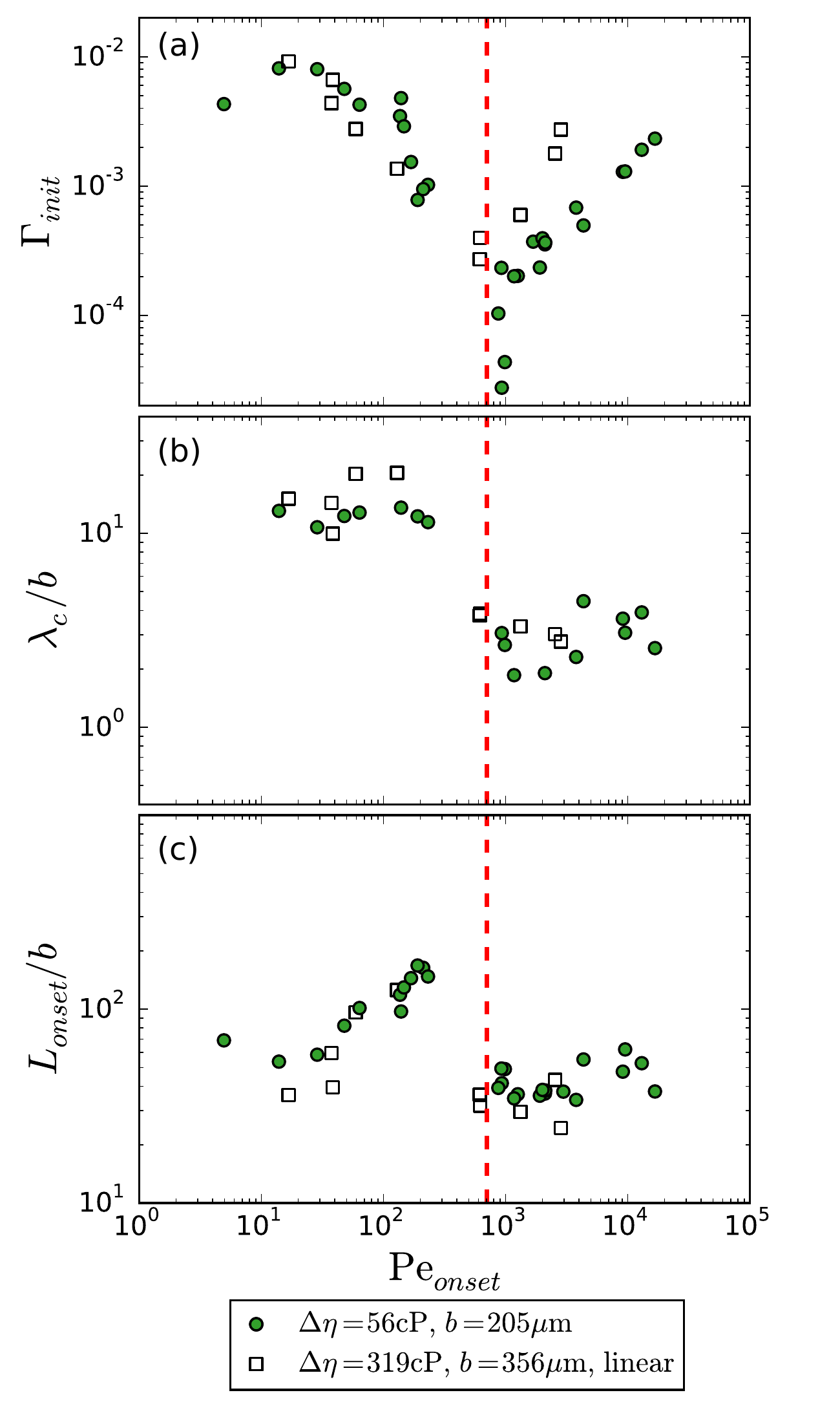}
	\caption{Comparison of data from the radial and linear cells with $\eta_{in}/\eta_{out} = 0.2$. (a) $\Gamma_{init}$, (b)  $\lambda_c/b$, and (c) $L_{onset}/b$, versus $\mathrm{Pe}_{onset}$. The red line denotes the same Pe number as in Fig.~\ref{fig:CombinedTransition_RfRi}. This shows quantitative agreement between the two geometries, except for a slight shift of Pe$^*$ for the linear cell.}
     \label{fig:CombinedTransition_lincomp}
\end{figure}

Figure~\ref{fig:CombinedTransition_RfRi} shows data as a function of $\mathrm{Pe}_{onset}$ at fixed viscosity ratio, $\eta_{in}/\eta_{out} = 0.2$ for a range of 
$b$, and viscosity differences, $\Delta \eta \equiv \eta_{out}-\eta_{in}$. Figure~\ref{fig:CombinedTransition_RfRi}a shows the finger growth rate at onset, $\Gamma_{init}$.   At $\mathrm{Pe}_{onset} = \mathrm{Pe}^* \approx 700$ there is a sharp feature where $\Gamma_{init}$ approaches zero.  As $\mathrm{Pe}^*$ is approached from either side, the growth rate drops precipitously.  At $\mathrm{Pe}^*$ both the normalized finger width, $\lambda_c/b$, and normalized onset length, $L_{onset}/b$, jump discontinuously, as shown in Figs.~\ref{fig:CombinedTransition_RfRi}b and c.  These data show a sharp transition between separate high- and low-Pe regimes with different morphological properties.  Figure~\ref{fig:CombinedTransition_lincomp} shows data from the linear cell compared to a set from the radial cell. Both geometries show the same behavior at the transition except that, for the linear cell, Pe$^*$ appears slightly shifted.

\begin{figure}
	\centering
	\includegraphics[width=6.2in]{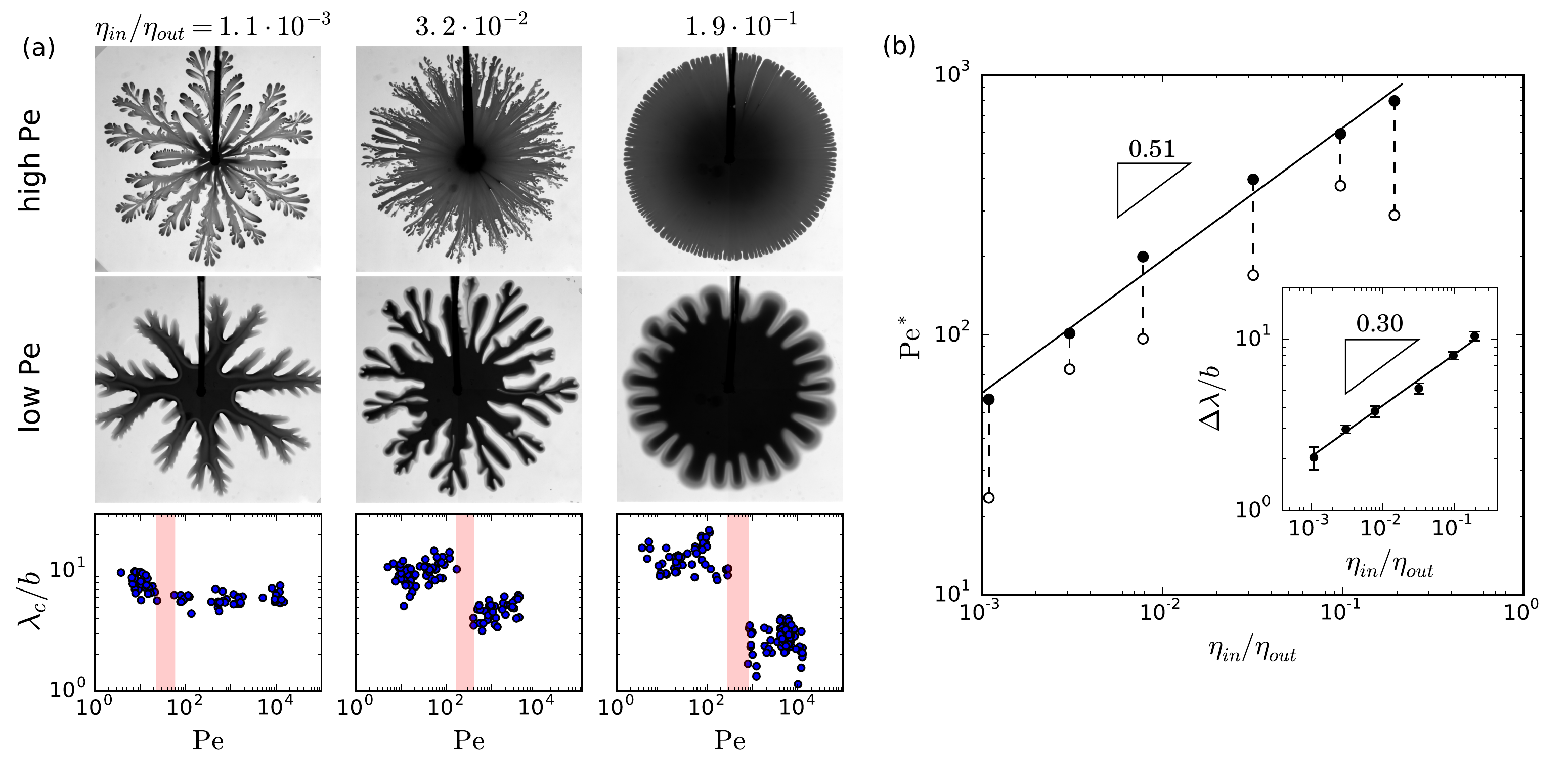}
	\caption{(a) Images from experiments in the high- and low-Pe regimes and corresponding measurements of $\lambda_c/b$ versus Pe for $\eta_{in}/\eta_{out}=1.1\cdot 10^{-3},\ 3.2\cdot 10^{-2},\ 1.9\cdot 10^{-1}$. Images taken when the outer radius of inner fluid is 5 cm. In the $\lambda_c/b$ data, all but the lowest $\eta_{in}/\eta_{out}$ show a clear jump in wavelength $\lambda_{c}/b$. The red shaded region shows the bounds on the critical Pe number, $\mathrm{Pe^*}$. (b) $\mathrm{Pe}^*$ versus $\eta_{in}/\eta_{out}$. Filled and open circles are the upper and lower bounds for $\mathrm{Pe^*}$ found in $\lambda_c/b$ data. Line shows power-law fit with exponent 0.51. Inset: magnitude of the jump in the unstable wavelength normalized by the plate spacing, $\Delta\lambda/b$, versus $\eta_{in}/\eta_{out}$. Line shows power-law fit with exponent 0.30.}
    \label{fig:CombinedEiEoData}
\end{figure}

The inset of Fig.~\ref{fig:CombinedTransition_RfRi}b shows measurements of $\lambda_c/b$ taken not at the point of onset but after the fingers have had a chance to grow; $\lambda_c/b$ does not vary appreciably as long as one remains in one or the other phase.  By measuring $\lambda_c/b$ one can tell what regime the system is in. In our radial cell with constant injection rate, the velocity of the interface decreases with the distance from the inlet.  Thus an experiment that begins on the high-Pe side of the transition, with the smaller wavelength fingers, will eventually exhibit fingers with characteristics of the low-Pe regime once the interfacial velocity has dropped sufficiently so that the interface has a Pe number below $\mathrm{Pe}^*$.

Figure~\ref{fig:CombinedEiEoData}a shows the patterns for experiments with viscosity ratios ranging from 0.001 to 0.2 in both the high- and low-Pe regimes. At low Pe the fingers are broader and more uniform in concentration than at high Pe indicating that the inner fluid fully fills the gap out to the broad interface. The onset length, $L_{onset}$, for the high-Pe regime decreases with $\eta_{in}/\eta_{out}$ and becomes comparable to the injection inlet for ratios smaller than 0.04 below which we are unable to measure quantities associated with the onset. 

Nevertheless, noting that $\lambda_c/b$ remains nearly constant within each regime, see plots in Fig.~\ref{fig:CombinedEiEoData}a, we can still measure the jump in its value, $\Delta\lambda_c/b$, between the two sides of the transition over the entire range of $\eta_{in}/\eta_{out}$ tested.  The inset of Fig.~\ref{fig:CombinedEiEoData}b, shows $\Delta\lambda_c/b\propto(\eta_{in}/\eta_{out})^{0.30 \pm 0.02}$.  By taking the lowest $\mathrm{Pe}$ at which we see a tip-splitting event for the high-$\mathrm{Pe}$ regime and the highest $\mathrm{Pe}$ at which we see diffusive fingering, we obtain bounds for $\mathrm{Pe}^*$.  Figure~\ref{fig:CombinedEiEoData}b shows $\mathrm{Pe}^* \propto (\eta_{in}/\eta_{out})^{0.51 \pm 0.03}$. We conclude that these high and low $\mathrm{Pe}$ regimes are robust features of the miscible fingering instability.

\begin{figure}
	\centering
	\includegraphics[width=2.7in]{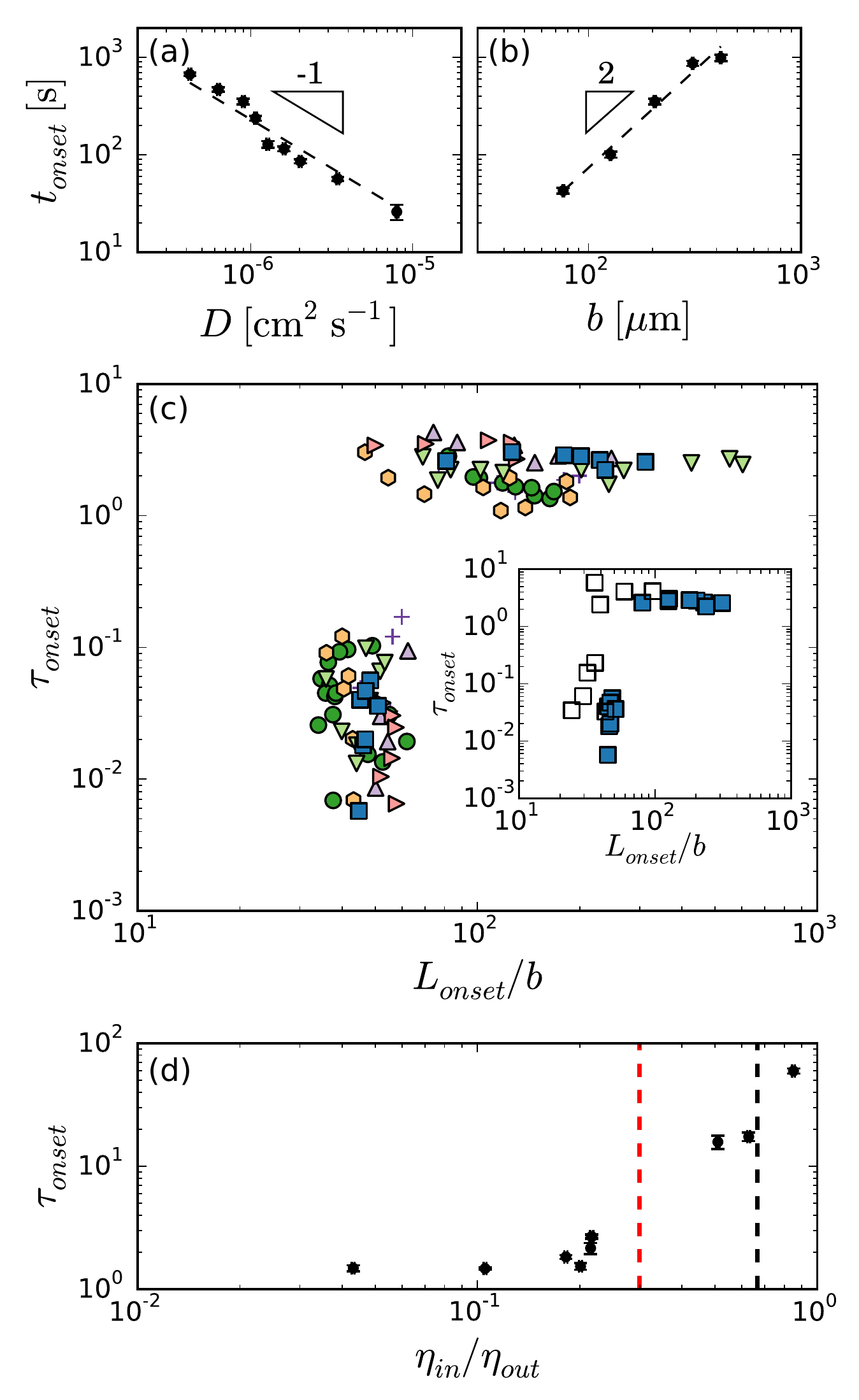}
	\caption{(a) $\tau_{onset}$ versus $D$ at $b=205\mu\mathrm{m}$. Dotted line shows $t_{onset} \propto D^{-1}$. (b) $\tau_{onset}$ versus $b$ with $D=0.90 \cdot 10^{-6} \mathrm{\ cm^2 s^{-1}}$. Dotted line shows $t_{onset} \propto b^2$.  (c) $\tau_{onset}$ versus $L_{onset}/b$. High-Pe regime has constant $L_{onset}/b$ while low-Pe regime has constant $\tau_{onset}$. Legend is the same as Fig.~\ref{fig:CombinedTransition_RfRi}. The inset shows a comparison between linear cell (open squares) and radial cell data.
	(d) $\tau_{onset}$ versus $\eta_{in}/\eta_{out}$. The red and black lines show stability thresholds from ref.~\citep{Bischofberger14} and ref.~\cite{Lajeunesse99} respectively.}
    \label{fig:Tonscaling}
\end{figure}

To gain insight into the re-emergence of fingering after the transient stability, we examine the instability onset as a function of time. Figure~\ref{fig:CombinedTransition_RfRi}c shows that at high $\mathrm{Pe}_{onset}$, the onset radius is constant while at low $\mathrm{Pe}_{onset}$ it is approximately proportional to $\mathrm{Pe}_{onset}$.  
If we plot the times of onset, $t_{onset}$, instead of $L_{onset}$, we find that for fluids with different diffusivities, $D$, and gap spacings, $b$, $t_{onset}\propto b^2/D$ as shown in Figs.~\ref{fig:Tonscaling}a and b. This suggests that $t_{onset}$ is related to a diffusive length: $\sqrt{4Dt}$. We therefore define a dimensionless onset time, $\tau_{onset} \equiv 4Dt_{onset}/b^2$.

The data for $\eta_{in}/\eta_{out}=0.2$, see Fig.~\ref{fig:Tonscaling}c, show that in the high-Pe regime, the onset is characterized by a constant value of the onset {\em length}, $L_{onset}/b$; in contrast, at low-Pe, the onset is characterized by a nearly constant value of the onset {\em time}, $\tau_{onset}$.  In Appendix D, we describe additional experiments showing that $\tau_{onset}$ is insensitive to the injection protocols. From this we conclude that it is the time rather than the length that robustly characterizes the onset in the low-Pe regime. Again, the linear cell shows consistent behavior compared to the radial cell, as shown in the inset of Fig.~\ref{fig:Tonscaling}c.

The Fig.~\ref{fig:Tonscaling}d shows $\tau_{onset}$ in the low-Pe regime versus viscosity ratio up to $\eta_{in}/\eta_{out} \approx 1.0$, which is above the cutoff reported for high-Pe fingering~\cite{Lajeunesse99,Bischofberger14}.  The cutoff has been ascribed~\cite{Lajeunesse99} to the structure at the front of the inner-fluid tongue changing from a sharp to a rounded profile.  In the low-Pe regime, these three-dimensional structures do not form, so that this additional region of stabilization disappears in accord with our results. For $\eta_{in}/\eta_{out} < 0.3$, $\tau_{onset} \sim O(1)$ indicates that the diffusion length is comparable to the gap spacing, $b$; however, above this threshold the value of $\tau_{onset}$ (and therefore the diffusion length) increases rapidly with increasing $\eta_{in}/\eta_{out}$.  In the absence of diffusion, this threshold, $\eta_{in}/\eta_{out} = 0.3$, is the viscosity ratio above which the profile of the inner-fluid profile no longer has a blunt tip (over a lengthscale $\sim b$) but becomes progressively thinner with increasing $\eta_{in}/\eta_{out}$. From this we conclude that, for large viscosity ratios, the diffusion length necessary to destabilize the interface becomes the length of the tapered finger not the distance between the plates, $b$.  That is, counter-intuitively, it is the longitudinal, not the transverse length scale in the gap which determines the appropriate amount of diffusion.

\begin{figure}
	\includegraphics[width=6in]{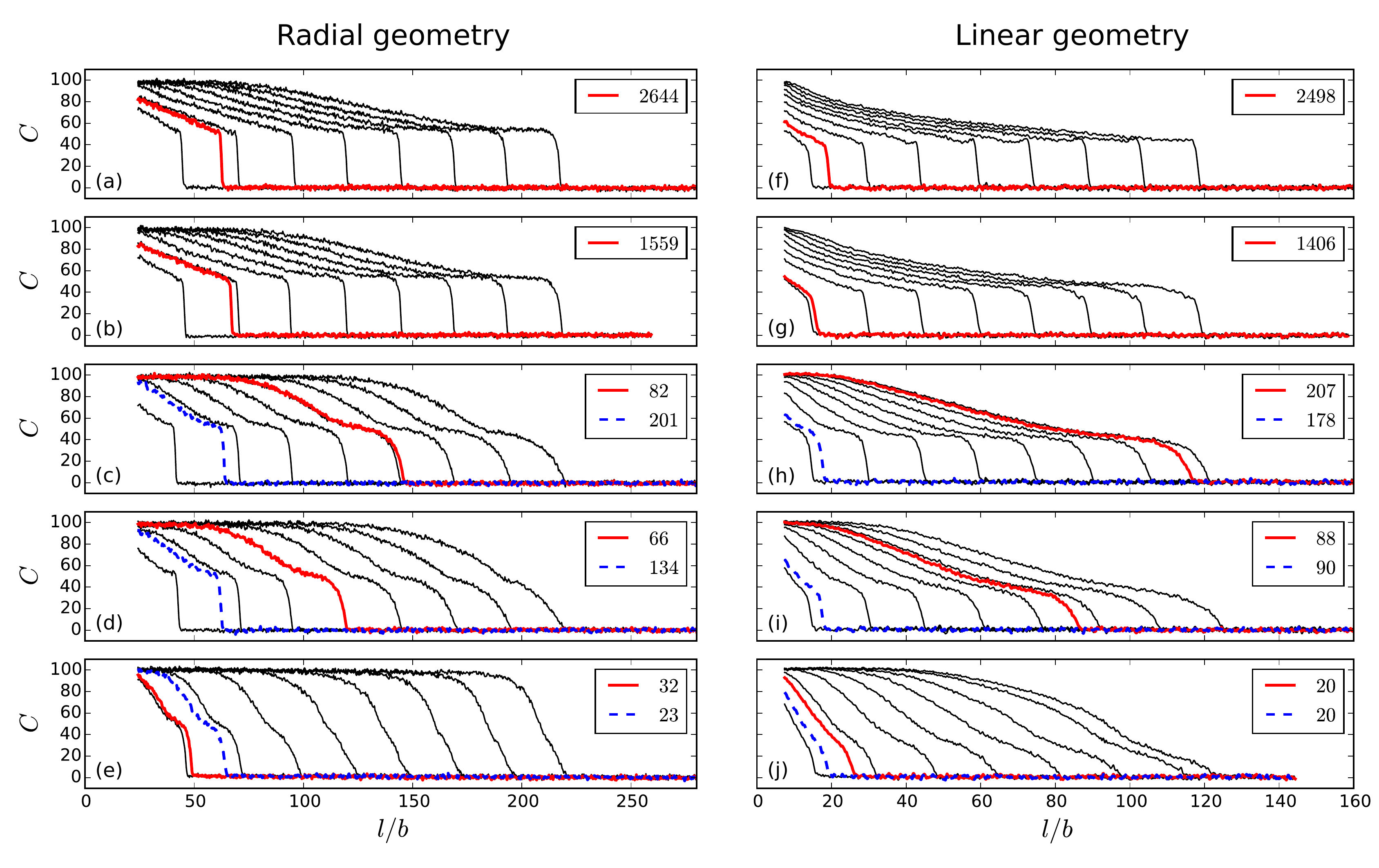}
	\caption{Concentration profiles spanning the transition between the two regimes of fingering. Profile on the left are from a radial geometry while those on the right are from a linear one.
	 In each set of curves the red (---) curve denotes the profile at the onset of fingering while the blue (- -) curve is the profile at the average onset length for the high-Pe regime. The labels for the red and blue curves are the Pe numbers for the corresponding profiles. (a), (b), (f), and (g) have fingers in the high-Pe regime while (c)-(e) and (h)-(j) are in the low-Pe regime.  Radial profiles here are taken with $b=205\mu$m and $\Delta\eta = 150$cP, linear profiles are taken with $b=356\mu$m and $\Delta\eta=319$cP. All profiles have $\eta_{in}/\eta_{out}=0.2$.}
    \label{fig:Concs}
\end{figure}

We now look at how the concentration profiles of the patterns change across the transition at Pe$^*$. Figure~\ref{fig:Concs} shows these profiles for experiments in both the radial and linear geometries. In the top two rows we show profiles from the high-Pe regime. The red curve denotes the profile at the onset of fingering for that experiment. Note that at onset, the tongue formation has not yet occurred; it is only after onset that a tongue, a flat protrusion, appears. We note that the high-Pe profiles look similar to what has been reported in previous work: the profile has a blunt tip and a tongue that fills roughly half of the gap~\cite{Lajeunesse97, Lajeunesse99, Bischofberger14, Yang97, Rakotomalala97, Talon13}. There are only slight differences between the radial and linear geometries that appear at the back of the finger near the inlet; the inner fluid fills the gap more in the radial cell than it does in the linear one.

For the low-Pe patterns, we see that even before the onset of fingering the inner fluid develops tongue structures similar to those seen in the high-Pe regime. Due to the increased effect of diffusion these tongues' concentration decreases with distance from the inlet. It is important to note that it is only at the lowest injection rates, after fingers have formed, that the concentration profiles begin to look like a diffusive front with no additional structure. Thus, it is only in this regime that purely two-dimensional theories that do not account for gap structure would be applicable~\cite{Chuoke59,Gardner84,Tan87,Yortsos87}. This suggests why our experiments do not show the scaling of $\lambda_c$ with Pe that those theories predict. 

If one looks at the low-Pe profiles at the same length as where onset occurs in the high-Pe regime, one sees subtle but very important differences. In the high-Pe regime the profile has a very sharp and abrupt front.  In the low-Pe regime, this profile is rounded -- with the degree of rounding increasing as Pe drops farther from Pe$^*$. If a blunt profile is indeed necessary for fingering to occur, then when diffusion rounds out this structure before the pattern can reach the onset length, the fingering would be prevented.  This hints at why diffusion helps to stabilize the patterns against fingering as Pe$^*$ is approached.

A final point to be made is that in Fig.~\ref{fig:IntroTransition}b there is a lighter grey region at the edge of the low-Pe regime patterns. This is seen in both radial and linear geometries. The evolution of the concentration profiles reveals that this is not a diffuse mixing region, but instead is the remnants of a tongue structure that forms at early times during injection. The formation of this tongue is a robust feature of how the inner fluid structure grows in the gap.  Because it also appears in the linear cell, it not due to a higher interface velocity near the inlet which would only apply to the radial cell.  We also note that the length of the tongue region in the onset profiles in the low-Pe regime (shown by the red curves in Fig.~\ref{fig:Concs}c-e and h-j), shrinks with lowering the Pe number; the onset of fingering in the low-Pe regime is independent of the length of the tongue.

\section{Discussion and Conclusion}

Our experiments on miscible pairs of fluids demonstrate that (i) at high $\mathrm{Pe}_{onset}$, three-dimensional structure within the gap is crucial for determining the nature of the fingering instability and (ii) at lower $\mathrm{Pe}_{onset}$, diffusion eradicates the three-dimensional structures so that the instability can profitably be considered as a two-dimensional problem.  The most dramatic aspect of these results is the well-defined transition between two distinct fingering regimes.  This transition appears to be continuous in one observable: the finger growth rate continuously decreases (approaching zero) at the critical $\mathrm{Pe}$ number and then re-emerges, apparently smoothly, at lower $\mathrm{Pe}_{onset}$. However there are other observables, such as $\lambda_c$ and $L_{onset}$, that appear to jump discontinuously at the transition. The existence of both smooth and discontinuous features at the transition is atypical and is reminiscent of a mixed-order phase transition, as has been seen at the jamming transition of spheres \cite{Liu2010}.
 
Because diffusion suppresses the miscible-fingering instability at small lengthscales, theories that treat the system as purely two-dimensional have predicted that $\lambda_c$ should depend on $\mathrm{Pe}$ \cite{Chuoke59,Gardner84,Tan87,Yortsos87}. In seeming contradiction, experiment (in the high-Pe regime) has found that $\lambda_c$ is insensitive to the injection rate or fluid properties and only depends on the gap spacing, $b$~\cite{Chen1987,Chen89,Bischofberger14,Lajeunesse97,Paterson85,Aubertin2009}.  However, recent simulations ~\cite{Goyal06,Oliveira11} that include the three-dimensional profile of the inner fluid have concluded that $\lambda_c$ is insensitive to $\mathrm{Pe}$ and viscosity ratio, in agreement with experiments.

Until now, there has been no comprehensive understanding of how these limiting cases, where the structure of the inner fluid is either two or three dimensional, are related to one another.  Our results showing a transition at $\mathrm{Pe}^*\gg 1$ (i.e., where advection dominates over diffusion) can reconcile the conflicting conclusions from experiment, theory and simulation.  They show that there is a regime at very low $\mathrm{Pe}$ where the inner fluid completely fills the gap; the two-dimensional picture is appropriate in this regime.  The conclusion that fingering persists for all $\eta_{in}/\eta_{out} < 1$ (as for immiscible systems) is corroborated by our experiments in the low-$\mathrm{Pe}$ regime.  At high $\mathrm{Pe}$, three-dimensional structure controls the transition; this is the regime that has hitherto been explored by experiments and simulations.  These are two separate regimes separated by a sharp transition; the system cannot be tuned continuously between them without encountering the transition and an accompanying disruption to the fingering.

A pronounced delay in the onset of fingering has also been observed in immiscible pairs of fluids~\cite{Ramachandran2017} as well as in the miscible fluids in both the high-$\mathrm{Pe}$ regime~\cite{Bischofberger14} and the low-$\mathrm{Pe}$ regime described here.  The cause of the delay in initiating the fingering patterns remains perplexing.  A geometrical argument for an onset radius based on the radial nature of the injection~\cite{Paterson81,Nagel13} estimates the onset radius to be very much smaller than the onset radii observed. Moreover, these arguments only use the fact that the velocity near the inlet is larger and would not be able to account for the delayed onset we have observed in the rectilinear cell. Clearly, further study of the fingering onset in all the different regimes and geometries is needed.  
  
Our work has demonstrated a novel transition in the viscous fingering instability most notably marked by the unexpected stability of the interface at $\mathrm{Pe}^*$, a critical value of $\mathrm{Pe}_{onset}$. The patterns in the regimes on the two sides of this transition show different morphologies, including differences in the most unstable wavelength and the profile structure of the inner fluid within the gap. The stabilization of the high-$\mathrm{Pe}$ regime at the transition is due to diffusion altering the profile of the fingers. The fingering in the low-$\mathrm{Pe}$ regime occurs after the three-dimensional structure is lost; in this regime, the system is quasi-two dimensional and no longer has the three-dimensional structure which had helped to stabilize the lateral patterns. These features highlight the importance of the gap structure and demonstrate a need for additional work at intermediate $\mathrm{Pe}$ to see how the structure within the gap dissolves and by what mechanism the inner fluid entirely fills the gap thereby ushering in the low-$\mathrm{Pe}$ regime. There is need for a modelling and theory effort to help provide a better understanding of these phenomena. These considerations also open up the possibility that, at even lower Pe, there may be a second transition to stability where diffusion acts not on the small lengthscale of the gap spacing, $b$, but on the larger scale of $\lambda_{c}$.

Previous techniques for controlling the viscous fingering instability, which can only work for immiscible fluids, exploit the stabilizing effect of surface tension~\cite{Li2009,AlHousseiny12,Pihler2012,Zheng2015}.  However, the discovery of a stable point in miscible fluids opens the possibility of halting the formation of fingers when no surface tension forces are present.  It provides a novel method for controlling fluid flow in miscible systems.

Due to the importance of the inner-fluid profile it is tantalizing to speculate if there are other ways to disrupt this structure and to observe their effects on pattern formation. Such experiments would allow a deeper understanding of how the onset of fingering occurs in the high-$\mathrm{Pe}$ regime.

\begin{acknowledgments}
We thank Irmgard Bischofberger, Rudro Rana Biswas, Todd Dupont, Paul Wiegmann and Tom Witten for useful discussions.  The work  was supported by the University of Chicago Materials Research Science and Engineering Center, which is funded by the National Science Foundation under award number DMR-1420709 and by NSF Grant DMR-1404841.
\end{acknowledgments}

\appendix

\section{Diffusion in a binary system}

For fluids that are binary mixtures of molecules, the diffusion is not characterized by a constant, but rather a coefficient that depends upon the local concentration of its components. In these systems a mutual diffusion coefficient, $D_{12}$, measures how the macroscopic concentration of the binary mixture diffuses due to gradients in the local mixture.  Because $D_{12}$ depends on concentration, a Fick's law is used to describe the changes in concentration in the following form:
\begin{equation}
\frac{\partial \phi}{\partial t} = \frac{\partial}{\partial x}\left( D_{12}(\phi)\frac{\partial\phi}{\partial x} \right)
\label{EQ:Fick}
\end{equation}
where $\phi$ denotes the molar fraction of the mixture.
At the extremes $\phi=0$ or $\phi=1$, $D_{12}$ should equal the self-diffusion of the two species. This is shown experimentally in Fig. 2 of D'Errico \textit{et al.} \cite{dErrico2004} in water-glycerol mixtures.

We can calculate an effective diffusion constant for our systems using the measured mutual diffusion between water-glycerol mixtures~\cite{dErrico2004}.  We need a single number, $D$, that approximates the diffusion between two initial concentrations of water and glycerol and that does not change during the course of a single experiment.

\begin{figure}
    \centering
	\includegraphics[width=3.4in]{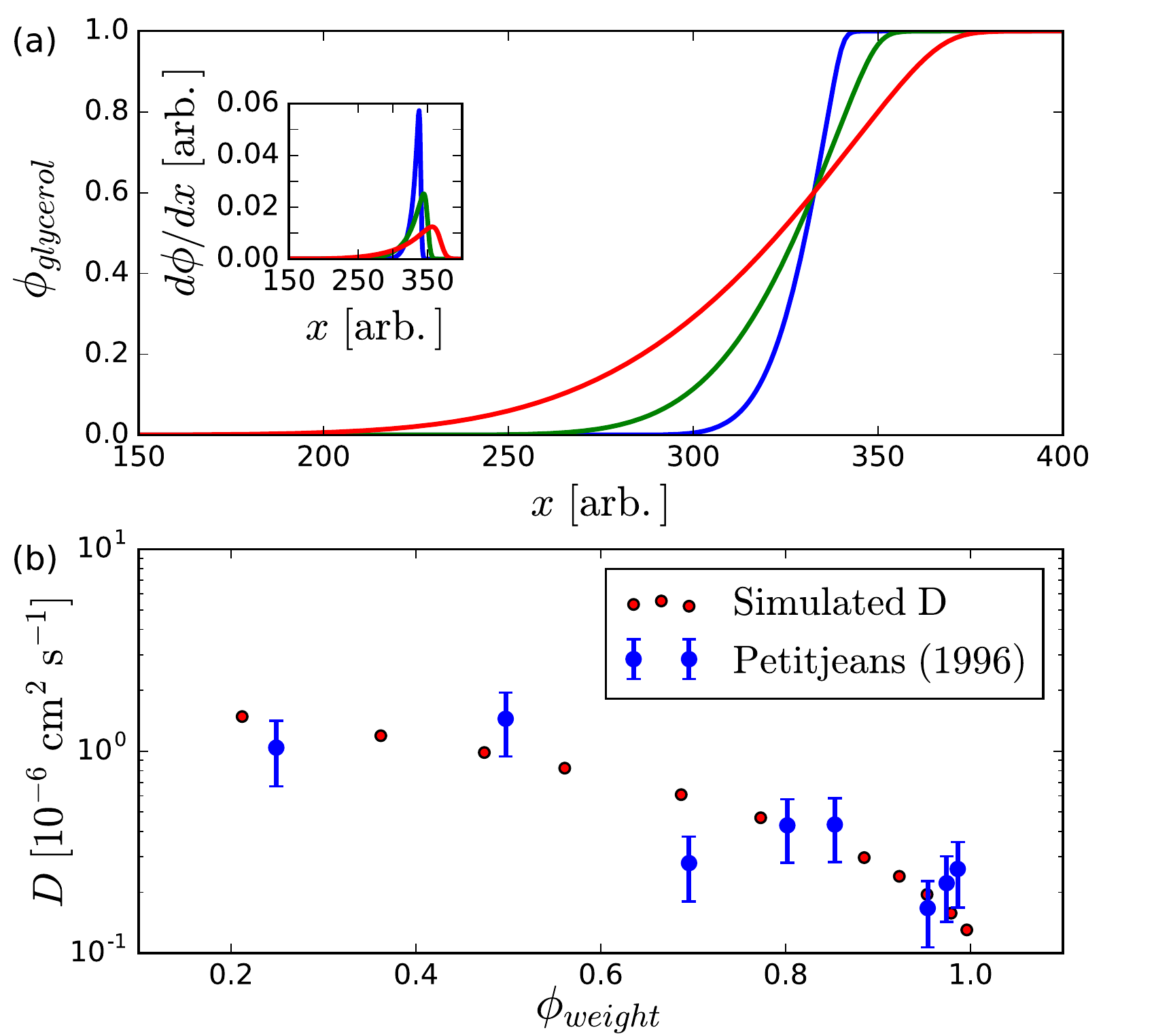}
	\caption{(a) The profiles of concentration, $\phi_{glycerol}$, at three different points in time from the numerics based on Eq. (\ref{EQ:Fick}). The inset shows the derivative of the concentration profile to highlight the asymmetry in the shape. (b) Plot of the effective diffusion constant versus the weight fraction of glycerol. The red circles are our simulated $D$ values and the blue circles are experimental values reported in \cite{Petitjeans1996}.}
	\label{Sim-comp}
\end{figure}

Using the modified Fick's law, Eq. (\ref{EQ:Fick}), an initial step-function concentration profile evolves to an asymmetric profile as shown in Fig.~\ref{Sim-comp}a. This is in contrast to a system with a constant diffusivity.  Petitjeans and Maxworthy~\cite{Petitjeans1996} have provided a method of approximating an effective diffusion constant, $D$, from the evolution of asymmetric profiles.  We follow that analysis here.  For a given spatial profile, $C(x)$, that has values of 0 and 1 for the initial concentrations of the two fluids, one can find the point, $X_{max}$, where the slope, $d^2C/dx^2=0$, is maximum. The difference between the outer-fluid concentration and the maximal-slope value is $\Delta C_g \equiv 1- C(X_{max})$; likewise $\Delta C_m \equiv C(X_{max})$ is the difference between the inner-fluid concentration and the maximal-slope value. One then determines where the concentration is equal to $C(X_{max})+(1-1/e)\Delta C_g$, and where the concentration is equal to $C(X_{max})-(1-1/e)\Delta C_m$ (see Fig. 4B in \cite{Petitjeans1996}). The distance between these two locations is defined as $\delta$. The effective diffusion constant, $D$, is obtained from $D=\delta^2/(6.35 t)$, where $t$ is the time taken to diffuse to a given concentration profile and the factor of 6.35 comes from carrying out this analysis on a system with a constant mutual diffusion coefficient.

We simulate Eq. (\ref{EQ:Fick}) using Euler's method to obtain values for our effective diffusion constants. The value of $D_{12}$ used is taken from Eq. (2) in \cite{dErrico2004}:
\begin{equation}
D_{12}\times 10^9\ \mathrm{m^2 s^{-1}} = \frac{(1.024\pm0.010)-(0.91\pm0.05)\phi}{1+(7.5\pm0.3)\phi}\pm0.004 .
\end{equation}
Our simulation is run with the concentration given in molar fraction and then the resulting profiles are converted to percent weight of glycerol. We then calculate $\delta$ for these concentrations using the prescription of Petitjeans and Maxworthy~\cite{Petitjeans1996}.

Petitjeans and Maxworthy took some experimental data on these effective diffusion constants between different initial concentration water-glycerol mixtures and pure glycerol~\cite{Petitjeans1996}.  Figure~\ref{Sim-comp}b compares our numerics to those measurements and find good agreement.

To simulate the diffusion for our experimental fluids we need to know the molar fraction of our mixtures. Reference \cite{Cheng2008} provides an empirical relation for the viscosity of water-glycerol mixtures based upon the temperature and weight concentration: 
\begin{equation}
\mu = \mu_w^{\alpha}\mu_g^{1-\alpha}
\end{equation}
where $\mu$, $\mu_w$, and $\mu_g$ are the viscosities for the mixture, pure water, and pure glycerol respectively. $\alpha$ is a weighting factor:
\begin{equation}
\alpha = 1 - \phi_m +\frac{ab\phi_m(1-\phi_m)}{a\phi_m+b(1-\phi_m)}
\end{equation}
where $\phi_m$ is the weight concentration of glycerol (1 being pure glycerol and 0 being pure water), and $a$ and $b$ depend on temperature. Since our experiments are conducted at a single temperature we take these numbers to be fitting constants. By using our measured values of viscosity at a fixed temperature we can use this formula to back out the molar fraction of our solutions. To account for a difference in temperature of our system we scale our results so that in the dilute limit of water as the inner solution diffusing into pure glycerol our diffusion constant is equal to the self-diffusion of water in a bath of glycerol, this scaling factor is about 0.943 for fluids at $22^{\circ}$C. It is slightly less than one since the values from \cite{dErrico2004} were measured at $25^{\circ}$C.

\section{Comparison of dyed images to schlieren imaging}

\begin{figure}
	\centering
	\includegraphics[width=3.4in]{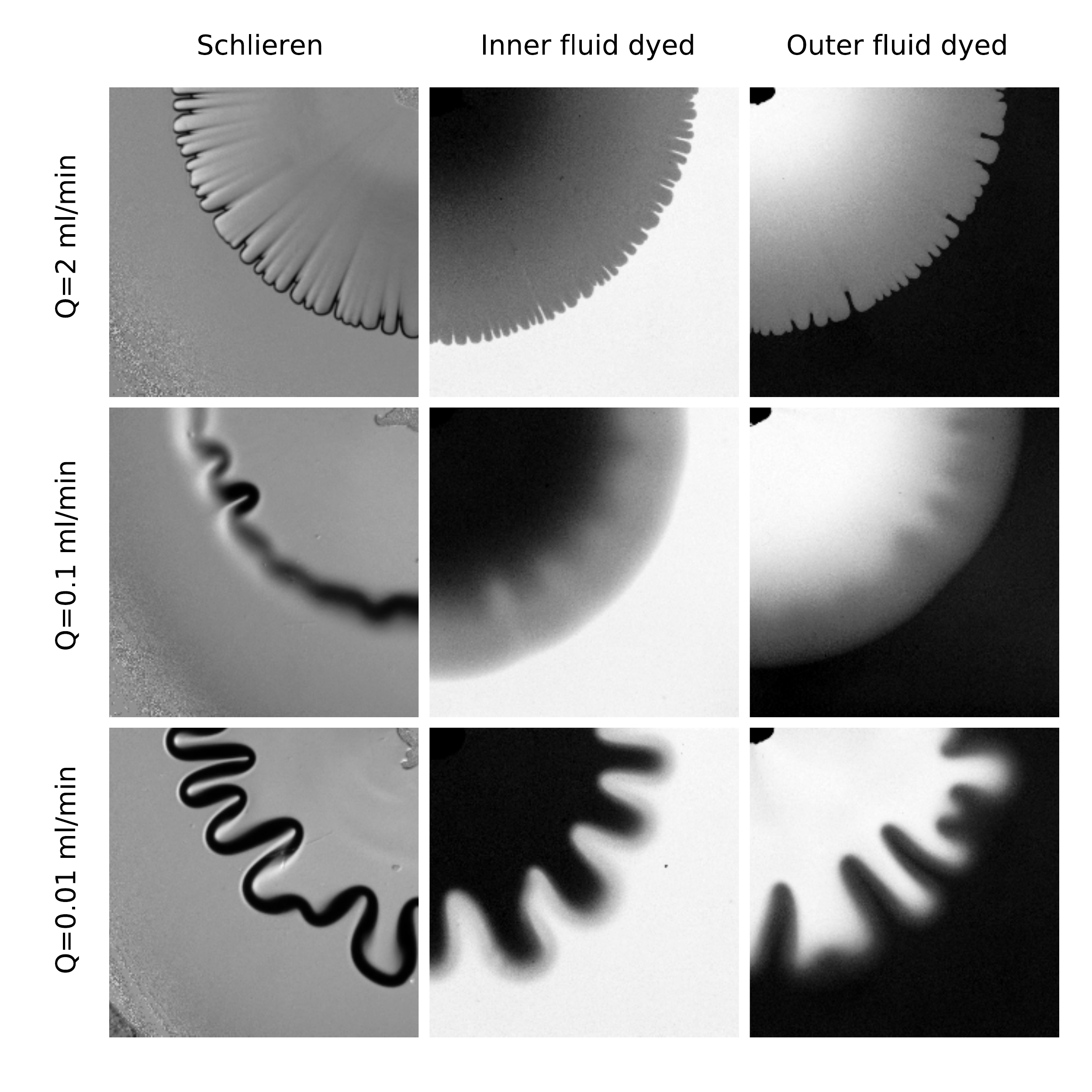}
	\caption{The images show different methods of visualizing the fluids in our experiments at different flow rates. The rows show, from top to bottom, flow rates of $2\ \mathrm{ml/min},\ 0.1\ \mathrm{ml/min},\ 0.01\ \mathrm{ml/min}$. From left to right, the columns depict final patterns seen with (i) schlieren imaging and standard imaging with (ii) inner and (iii) outer fluid dyed. The schlieren images have been divided by the background to account for non-uniform lighting. Dyed images were taken when the outer extent of the dyed patterns reach a radius of 3.5 cm, the schlieren images were taken so that the total time of injection matches that of the corresponding experiments with dyed fluids. A final radius is not considered for schlieren since the diffuse boundary is not clearly visible for lower injection rates.  All experiments were done with $b=205\mu\mathrm{m}$, $\eta_{in}=24.7\ \mathrm{cP}$ and $\eta_{out}=116\ \mathrm{cP}$.}
	\label{Img-comp}
\end{figure}

Here we show comparisons between different types of imaging over a range of injection rates, Q.  Figure~\ref{Img-comp} shows images of fingering patterns from a schlieren optics setup and from back-lit photography of fluids with either dyed inner or outer fluid.  There are slight differences between the schlieren and the dyed inner-fluid images.  Schlieren imaging picks up gradients of index of refraction corresponding to gradients in concentration.  At high injection rates the images match nicely: both show a region with no change in concentration (near the inlet) and then a slow decrease in concentration out to the tips of fingers. At intermediate injection rates, the schlieren imaging is not sensitive to the outer most edge of the pattern. However, all techniques show the emergence of thicker fingers in the interior of the pattern at the same radial distance from the inlet. The size of this transition region between fully filled and the diffuse boundary is larger in the dyed image.  Finally, the images match well for the lowest injection rate. The schlieren images show no appreciable signal in the interior of the pattern.  This confirms the observation that the inner fluid has a constant thickness out to the edge of the pattern. Since both methods capture the same essential aspects of the patterns, it confirms that those observations are not artifacts from using dye in our fluids.  Finally, there is no appreciable difference at any of the injection rates between the two types of dyed images (the middle and right columns of the figure).

\section{Gravitational Effects}

Following the definition of $F$ from the main text, here we compute its largest value for each set of data from the radial cell. Since we report Pe number instead of $V$ we substitute $V=\mathrm{Pe} D/b$. For each value of b used with each fluid pair and the lowest Pe number we have measured, these provide values of $F$ found in Table~\ref{tab:Fnumber}. From this it is clear that even for the fluids with larger density differences viscous forces dominate. We also note that across our glycerol-water experiments there is an order of magnitude change in density differences with no quantitative change in the transition behavior. This leads us to conclude that gravitational forces can be neglected.



\begin{table}
\centering

\begin{tabular}{ c | c | c | c | c | c | c }
$\eta_{in}/\eta_{out}$ & $\Delta\eta\ [\mathrm{cP}]$ & $\Delta\rho\ [\mathrm{kg m^{-3}}]$ & $D \cdot 10^{-6}\ [\mathrm{cm^2 s^{-1}}]$ & $b\ [\mu\mathrm{m}]$ & $Pe_{lowest}$ & $F_{highest} \times 10^{-4}$ \\
\hline
0.20	&3.6	&128	&7.99	&205	&100	&30.1 \\
\hline
0.18	&20.2	&69.4	&0.345	&205	&49.8	&13.8 \\
\hline
0.22	&56.5	&43.3	&2.03	&419	&47.3	&45.3 \\
	&	&	&	&205	&4.94	&50.7 \\
	&	&	&	&127	&59.6	&0.999 \\
	\hline
0.20	&147	&35.3	&1.27	&205	&30.5	&4.17 \\
\hline
0.22	&256	&28.8	&0.900	&308	&15.6	&18.0 \\
	&	&   &	&205	&30	    &2.76 \\
	&	&	&	&127	&64.1	&0.308 \\
	&	&	&	&76	    &48.1	&0.0878 \\
	\hline
0.22	&687	&23.1	&0.421	&205	&74.8	&0.708 \\
\hline
0.097	&187	&50.8	&0.145	&205	&19.9	&0.0718 \\
\hline
0.032	&697	&68.6	&0.0942	&205	&5.05	&0.169 \\
\hline
0.0078	&546	&129	&0.158	&205	&5.81	&0.216 \\
\hline
0.0031	&325	&236	&0.314	&205	&1.93	&1.01 \\
\hline
0.0011	&1070	&251	&0.256	&205	&3.73	&0.207 \\
\hline
0.21 &319 &26.8 &1.09  &356 &16.7 & 16.1\\

\end{tabular}
\caption{Experimental fluid parameters and highest graviational number. The last line is for the fluid used in the linear cell experiments.}
\label{tab:Fnumber}
\end{table}

\section{Onset time dependence on injection rate schemes}


The  low-Pe regime occurs when the inner-fluid profile becomes uniform along the entire length of a finger and fully fills the gap.  However, the inner fluid tongue fully filling the gap cannot be due to passive diffusion by itself since diffusion would act to spread out the concentration 
and not act to increase the local concentration. Thus  some advective motion to increase the thickness of the fluid tongue must be at play.

\begin{figure}
	\centering
	\includegraphics[width=3.4in]{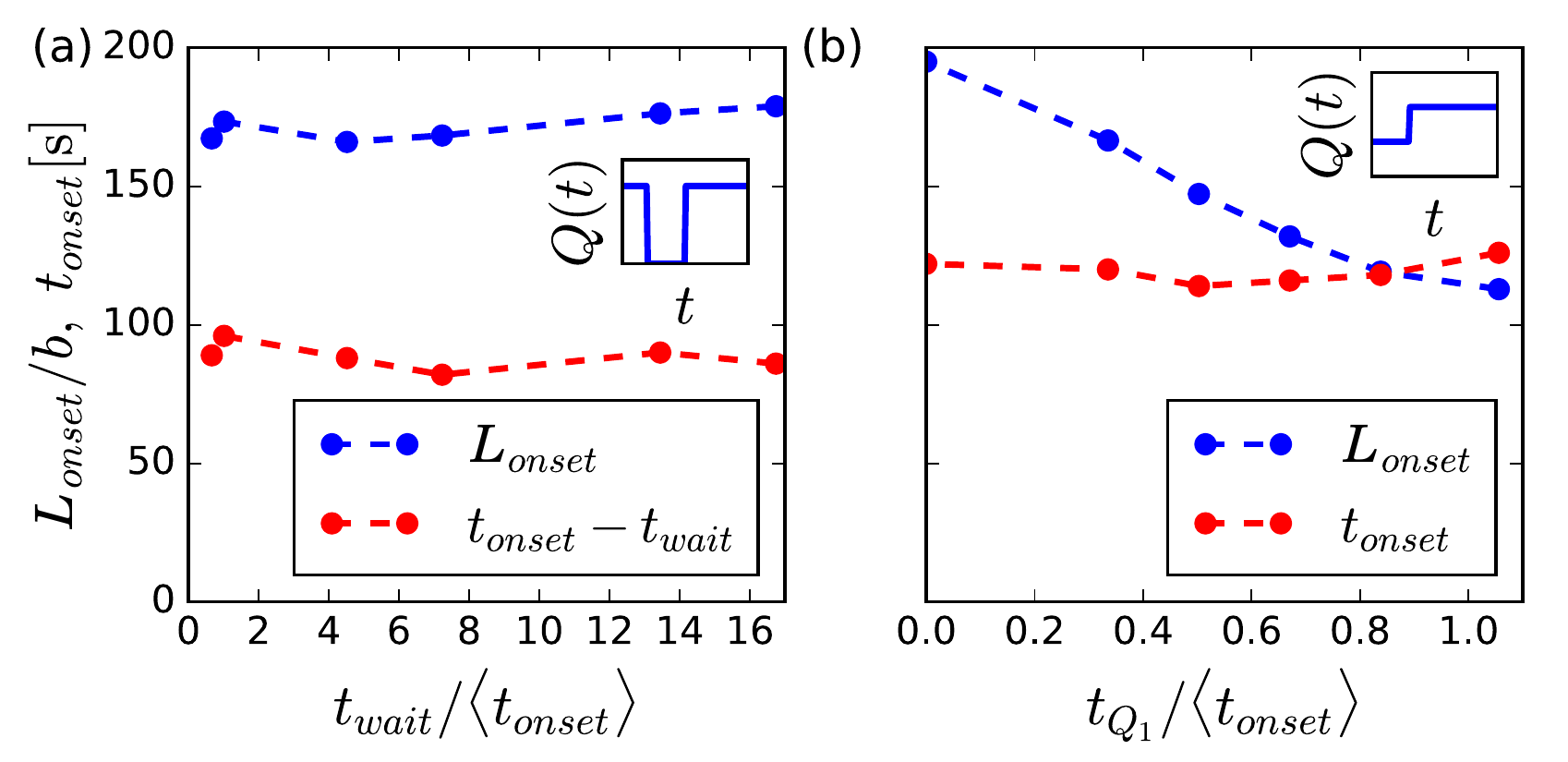}
	\caption{$L_{onset}$ (blue) and $t_{onset}$ (red), are shown for radial cell experiments with different time-varying injection rates $Q(t)$, shown schematically in the insets. The average onset time, $\langle t_{onset}\rangle$, is taken from experiments that use a single, non-varying injection rate. (a) After injection at a constant rate to a radius of 3 cm, injection stops for a time $t_{wait}$ before continuing. (b) The injection rate ($Q$) is low for a time  $t_{Q_1}$ before $Q$ is increased.}
	\label{fig:VariableQ}
\end{figure}

To show that passive diffusion alone cannot initiate the low-Pe regime, we conduct a series of experiments in the radial cell where the inner fluid is injected for a time shorter than the measured onset time and then let the fluid remain at rest for a time $t_{wait}$, after which we resume injection. Figure~\ref{fig:VariableQ}a shows the onset length and the time of onset (neglecting the time spent waiting) remain unchanged, even for waiting times over ten times longer than the expected onset time. We conclude that advection and diffusion coupled together are important for filling the gap.

We also vary the injection by first having a low volumetric rate and then, as shown schematically in the inset to Fig.~\ref{fig:VariableQ}b, by increasing the injection rate. Figure~\ref{fig:VariableQ}b shows that that the onset time remains constant, while the onset length decreases as the time that is spent injecting at the low rate is varied. When the injection rate is increased linearly with time, the onset time is unchanged as well. This implies that a constant $\tau_{onset}$ is a robust feature of the low-Pe regime.

\bibliography{diffusive-fingering.bib}

\begin{thebibliography}{42}%
\makeatletter
\providecommand \@ifxundefined [1]{%
 \@ifx{#1\undefined}
}%
\providecommand \@ifnum [1]{%
 \ifnum #1\expandafter \@firstoftwo
 \else \expandafter \@secondoftwo
 \fi
}%
\providecommand \@ifx [1]{%
 \ifx #1\expandafter \@firstoftwo
 \else \expandafter \@secondoftwo
 \fi
}%
\providecommand \natexlab [1]{#1}%
\providecommand \enquote  [1]{``#1''}%
\providecommand \bibnamefont  [1]{#1}%
\providecommand \bibfnamefont [1]{#1}%
\providecommand \citenamefont [1]{#1}%
\providecommand \href@noop [0]{\@secondoftwo}%
\providecommand \href [0]{\begingroup \@sanitize@url \@href}%
\providecommand \@href[1]{\@@startlink{#1}\@@href}%
\providecommand \@@href[1]{\endgroup#1\@@endlink}%
\providecommand \@sanitize@url [0]{\catcode `\\12\catcode `\$12\catcode
  `\&12\catcode `\#12\catcode `\^12\catcode `\_12\catcode `\%12\relax}%
\providecommand \@@startlink[1]{}%
\providecommand \@@endlink[0]{}%
\providecommand \url  [0]{\begingroup\@sanitize@url \@url }%
\providecommand \@url [1]{\endgroup\@href {#1}{\urlprefix }}%
\providecommand \urlprefix  [0]{URL }%
\providecommand \Eprint [0]{\href }%
\providecommand \doibase [0]{http://dx.doi.org/}%
\providecommand \selectlanguage [0]{\@gobble}%
\providecommand \bibinfo  [0]{\@secondoftwo}%
\providecommand \bibfield  [0]{\@secondoftwo}%
\providecommand \translation [1]{[#1]}%
\providecommand \BibitemOpen [0]{}%
\providecommand \bibitemStop [0]{}%
\providecommand \bibitemNoStop [0]{.\EOS\space}%
\providecommand \EOS [0]{\spacefactor3000\relax}%
\providecommand \BibitemShut  [1]{\csname bibitem#1\endcsname}%
\let\auto@bib@innerbib\@empty
\bibitem [{\citenamefont {Turing}(1952)}]{Turing1952}%
  \BibitemOpen
  \bibfield  {author} {\bibinfo {author} {\bibfnamefont {A.~M.}\ \bibnamefont
  {Turing}},\ }\bibfield  {title} {\enquote {\bibinfo {title} {The chemical
  basis of morphogenesis},}\ }\href@noop {} {\bibfield  {journal} {\bibinfo
  {journal} {Philos. Trans. Royal Soc. B}\ }\textbf {\bibinfo {volume} {237}},\
  \bibinfo {pages} {37--72} (\bibinfo {year} {1952})}\BibitemShut {NoStop}%
\bibitem [{\citenamefont {Kondo}\ and\ \citenamefont
  {Miura}(2010)}]{Kondo2010}%
  \BibitemOpen
  \bibfield  {author} {\bibinfo {author} {\bibfnamefont {S.}~\bibnamefont
  {Kondo}}\ and\ \bibinfo {author} {\bibfnamefont {T.}~\bibnamefont {Miura}},\
  }\bibfield  {title} {\enquote {\bibinfo {title} {Reaction-diffusion model as
  a framework for understanding biological pattern formation},}\ }\href@noop {}
  {\bibfield  {journal} {\bibinfo  {journal} {Science}\ }\textbf {\bibinfo
  {volume} {329}},\ \bibinfo {pages} {1616--1620} (\bibinfo {year}
  {2010})}\BibitemShut {NoStop}%
\bibitem [{\citenamefont {Bensimon}\ \emph {et~al.}(1986)\citenamefont
  {Bensimon}, \citenamefont {Kadanoff}, \citenamefont {Liang}, \citenamefont
  {Shraiman},\ and\ \citenamefont {Tang}}]{Bensimon86}%
  \BibitemOpen
  \bibfield  {author} {\bibinfo {author} {\bibfnamefont {D.}~\bibnamefont
  {Bensimon}}, \bibinfo {author} {\bibfnamefont {L.~P.}\ \bibnamefont
  {Kadanoff}}, \bibinfo {author} {\bibfnamefont {S.}~\bibnamefont {Liang}},
  \bibinfo {author} {\bibfnamefont {B.~I.}\ \bibnamefont {Shraiman}}, \ and\
  \bibinfo {author} {\bibfnamefont {C.}~\bibnamefont {Tang}},\ }\bibfield
  {title} {\enquote {\bibinfo {title} {Viscous flows in two dimensions},}\
  }\href@noop {} {\bibfield  {journal} {\bibinfo  {journal} {Rev. Mod. Phys.}\
  }\textbf {\bibinfo {volume} {58}},\ \bibinfo {pages} {977--999} (\bibinfo
  {year} {1986})}\BibitemShut {NoStop}%
\bibitem [{\citenamefont {Homsy}(1987)}]{Homsy87}%
  \BibitemOpen
  \bibfield  {author} {\bibinfo {author} {\bibfnamefont {G.~M.}\ \bibnamefont
  {Homsy}},\ }\bibfield  {title} {\enquote {\bibinfo {title} {Viscous fingering
  in porous media},}\ }\href@noop {} {\bibfield  {journal} {\bibinfo  {journal}
  {Annu. Rev. Fluid Mech.}\ }\textbf {\bibinfo {volume} {19}},\ \bibinfo
  {pages} {271--311} (\bibinfo {year} {1987})}\BibitemShut {NoStop}%
\bibitem [{\citenamefont {Saffman}\ and\ \citenamefont
  {Taylor}(1958)}]{Saffman58}%
  \BibitemOpen
  \bibfield  {author} {\bibinfo {author} {\bibfnamefont {P.~G.}\ \bibnamefont
  {Saffman}}\ and\ \bibinfo {author} {\bibfnamefont {G.~I.}\ \bibnamefont
  {Taylor}},\ }\bibfield  {title} {\enquote {\bibinfo {title} {The penetration
  of a fluid into a porous medium or hele-shaw cell containing a more viscous
  liquid},}\ }\href@noop {} {\bibfield  {journal} {\bibinfo  {journal} {Proc.
  Royal Soc. London Ser. A}\ }\textbf {\bibinfo {volume} {245}},\ \bibinfo
  {pages} {312--329} (\bibinfo {year} {1958})}\BibitemShut {NoStop}%
\bibitem [{\citenamefont {Nagel}(2017)}]{Nagel2017}%
  \BibitemOpen
  \bibfield  {author} {\bibinfo {author} {\bibfnamefont {S.~R.}\ \bibnamefont
  {Nagel}},\ }\bibfield  {title} {\enquote {\bibinfo {title} {Experimental
  soft-matter science},}\ }\href@noop {} {\bibfield  {journal} {\bibinfo
  {journal} {Rev. Mod. Phys.}\ }\textbf {\bibinfo {volume} {89}} (\bibinfo
  {year} {2017})}\BibitemShut {NoStop}%
\bibitem [{\citenamefont {Witten~Jr}\ and\ \citenamefont
  {Sander}(1981)}]{Witten1981}%
  \BibitemOpen
  \bibfield  {author} {\bibinfo {author} {\bibfnamefont {T.A.}\ \bibnamefont
  {Witten~Jr}}\ and\ \bibinfo {author} {\bibfnamefont {L.~M.}\ \bibnamefont
  {Sander}},\ }\bibfield  {title} {\enquote {\bibinfo {title}
  {Diffusion-limited aggregation, a kinetic critical phenomenon},}\ }\href@noop
  {} {\bibfield  {journal} {\bibinfo  {journal} {Phys. Rev. Lett.}\ }\textbf
  {\bibinfo {volume} {47}},\ \bibinfo {pages} {1400--1403} (\bibinfo {year}
  {1981})}\BibitemShut {NoStop}%
\bibitem [{\citenamefont {Paterson}(1984)}]{Paterson84DLA}%
  \BibitemOpen
  \bibfield  {author} {\bibinfo {author} {\bibfnamefont {L.}~\bibnamefont
  {Paterson}},\ }\bibfield  {title} {\enquote {\bibinfo {title}
  {Diffusion-limited aggregation and two-fluid displacements in porous
  media},}\ }\href@noop {} {\bibfield  {journal} {\bibinfo  {journal} {Phys.
  Rev. Lett.}\ }\textbf {\bibinfo {volume} {52}},\ \bibinfo {pages} {1621}
  (\bibinfo {year} {1984})}\BibitemShut {NoStop}%
\bibitem [{\citenamefont {Sander}\ \emph {et~al.}(1985)\citenamefont {Sander},
  \citenamefont {Ramanlal},\ and\ \citenamefont {Ben-Jacob}}]{Sander85}%
  \BibitemOpen
  \bibfield  {author} {\bibinfo {author} {\bibfnamefont {L.~M.}\ \bibnamefont
  {Sander}}, \bibinfo {author} {\bibfnamefont {P.}~\bibnamefont {Ramanlal}}, \
  and\ \bibinfo {author} {\bibfnamefont {E.}~\bibnamefont {Ben-Jacob}},\
  }\bibfield  {title} {\enquote {\bibinfo {title} {Diffusion-limited
  aggregation as a deterministic growth process},}\ }\href@noop {} {\bibfield
  {journal} {\bibinfo  {journal} {Phys. Rev. A}\ }\textbf {\bibinfo {volume}
  {32}},\ \bibinfo {pages} {3160} (\bibinfo {year} {1985})}\BibitemShut
  {NoStop}%
\bibitem [{\citenamefont {Nittmann}\ \emph {et~al.}(1985)\citenamefont
  {Nittmann}, \citenamefont {Daccord},\ and\ \citenamefont
  {Stanley}}]{Nittmann1985}%
  \BibitemOpen
  \bibfield  {author} {\bibinfo {author} {\bibfnamefont {J.}~\bibnamefont
  {Nittmann}}, \bibinfo {author} {\bibfnamefont {G.}~\bibnamefont {Daccord}}, \
  and\ \bibinfo {author} {\bibfnamefont {H.~E.}\ \bibnamefont {Stanley}},\
  }\bibfield  {title} {\enquote {\bibinfo {title} {Fractal growth viscous
  fingers: quantitative characterization of a fluid instability phenomenon},}\
  }\href@noop {} {\bibfield  {journal} {\bibinfo  {journal} {Nature}\ }\textbf
  {\bibinfo {volume} {314}},\ \bibinfo {pages} {141--144} (\bibinfo {year}
  {1985})}\BibitemShut {NoStop}%
\bibitem [{\citenamefont {Sander}(1986)}]{Sander86}%
  \BibitemOpen
  \bibfield  {author} {\bibinfo {author} {\bibfnamefont {L.~M.}\ \bibnamefont
  {Sander}},\ }\bibfield  {title} {\enquote {\bibinfo {title} {Fractal growth
  processes},}\ }\href@noop {} {\bibfield  {journal} {\bibinfo  {journal}
  {Nature}\ }\textbf {\bibinfo {volume} {322}},\ \bibinfo {pages} {789--793}
  (\bibinfo {year} {1986})}\BibitemShut {NoStop}%
\bibitem [{\citenamefont {Bettelheim}\ \emph {et~al.}(2005)\citenamefont
  {Bettelheim}, \citenamefont {Agam}, \citenamefont {Zabrodin},\ and\
  \citenamefont {Wiegmann}}]{Wiegmann05}%
  \BibitemOpen
  \bibfield  {author} {\bibinfo {author} {\bibfnamefont {E.}~\bibnamefont
  {Bettelheim}}, \bibinfo {author} {\bibfnamefont {O.}~\bibnamefont {Agam}},
  \bibinfo {author} {\bibfnamefont {A.}~\bibnamefont {Zabrodin}}, \ and\
  \bibinfo {author} {\bibfnamefont {P.}~\bibnamefont {Wiegmann}},\ }\bibfield
  {title} {\enquote {\bibinfo {title} {Singularities of the hele-shaw flow and
  shock waves in dispersive media},}\ }\href {\doibase
  10.1103/PhysRevLett.95.244504} {\bibfield  {journal} {\bibinfo  {journal}
  {Phys. Rev. Lett.}\ }\textbf {\bibinfo {volume} {95}},\ \bibinfo {pages}
  {244504} (\bibinfo {year} {2005})}\BibitemShut {NoStop}%
\bibitem [{\citenamefont {Praud}\ and\ \citenamefont
  {Swinney}(2005)}]{Swinney05}%
  \BibitemOpen
  \bibfield  {author} {\bibinfo {author} {\bibfnamefont {O.}~\bibnamefont
  {Praud}}\ and\ \bibinfo {author} {\bibfnamefont {H.~L.}\ \bibnamefont
  {Swinney}},\ }\bibfield  {title} {\enquote {\bibinfo {title} {Fractal
  dimension and unscreened angles measured for radial viscous fingering},}\
  }\href {\doibase 10.1103/PhysRevE.72.011406} {\bibfield  {journal} {\bibinfo
  {journal} {Phys. Rev. E}\ }\textbf {\bibinfo {volume} {72}},\ \bibinfo
  {pages} {011406} (\bibinfo {year} {2005})}\BibitemShut {NoStop}%
\bibitem [{\citenamefont {Cheng}\ \emph {et~al.}(2008)\citenamefont {Cheng},
  \citenamefont {Xu}, \citenamefont {Patterson}, \citenamefont {Jaeger},\ and\
  \citenamefont {Nagel}}]{Cheng08}%
  \BibitemOpen
  \bibfield  {author} {\bibinfo {author} {\bibfnamefont {X.}~\bibnamefont
  {Cheng}}, \bibinfo {author} {\bibfnamefont {L.}~\bibnamefont {Xu}}, \bibinfo
  {author} {\bibfnamefont {A.}~\bibnamefont {Patterson}}, \bibinfo {author}
  {\bibfnamefont {H.~M.}\ \bibnamefont {Jaeger}}, \ and\ \bibinfo {author}
  {\bibfnamefont {S.~R.}\ \bibnamefont {Nagel}},\ }\bibfield  {title} {\enquote
  {\bibinfo {title} {Towards the zero-surface-tension limit in granular
  fingering instability},}\ }\href@noop {} {\bibfield  {journal} {\bibinfo
  {journal} {Nat. Phys.}\ }\textbf {\bibinfo {volume} {4}},\ \bibinfo {pages}
  {234--237} (\bibinfo {year} {2008})}\BibitemShut {NoStop}%
\bibitem [{\citenamefont {Lajeunesse}\ \emph {et~al.}(1997)\citenamefont
  {Lajeunesse}, \citenamefont {Martin}, \citenamefont {Rakotomalala},\ and\
  \citenamefont {Salin}}]{Lajeunesse97}%
  \BibitemOpen
  \bibfield  {author} {\bibinfo {author} {\bibfnamefont {E.}~\bibnamefont
  {Lajeunesse}}, \bibinfo {author} {\bibfnamefont {J.}~\bibnamefont {Martin}},
  \bibinfo {author} {\bibfnamefont {N.}~\bibnamefont {Rakotomalala}}, \ and\
  \bibinfo {author} {\bibfnamefont {D.}~\bibnamefont {Salin}},\ }\bibfield
  {title} {\enquote {\bibinfo {title} {3d instability of miscible displacements
  in a hele-shaw cell},}\ }\href@noop {} {\bibfield  {journal} {\bibinfo
  {journal} {Phys. Rev. Lett.}\ }\textbf {\bibinfo {volume} {79}},\ \bibinfo
  {pages} {5254--5257} (\bibinfo {year} {1997})}\BibitemShut {NoStop}%
\bibitem [{\citenamefont {Lajeunesse}\ \emph {et~al.}(1999)\citenamefont
  {Lajeunesse}, \citenamefont {Martin}, \citenamefont {Rakotomalala},
  \citenamefont {Salin},\ and\ \citenamefont {Yortsos}}]{Lajeunesse99}%
  \BibitemOpen
  \bibfield  {author} {\bibinfo {author} {\bibfnamefont {E.}~\bibnamefont
  {Lajeunesse}}, \bibinfo {author} {\bibfnamefont {J.}~\bibnamefont {Martin}},
  \bibinfo {author} {\bibfnamefont {N.}~\bibnamefont {Rakotomalala}}, \bibinfo
  {author} {\bibfnamefont {D.}~\bibnamefont {Salin}}, \ and\ \bibinfo {author}
  {\bibfnamefont {Y.~C.}\ \bibnamefont {Yortsos}},\ }\bibfield  {title}
  {\enquote {\bibinfo {title} {Miscible displacement in a hele-shaw cell at
  high rates},}\ }\href@noop {} {\bibfield  {journal} {\bibinfo  {journal} {J.
  Fluid Mech.}\ }\textbf {\bibinfo {volume} {398}},\ \bibinfo {pages}
  {299--319} (\bibinfo {year} {1999})}\BibitemShut {NoStop}%
\bibitem [{\citenamefont {Bischofberger}\ \emph {et~al.}(2014)\citenamefont
  {Bischofberger}, \citenamefont {Ramachandran},\ and\ \citenamefont
  {Nagel}}]{Bischofberger14}%
  \BibitemOpen
  \bibfield  {author} {\bibinfo {author} {\bibfnamefont {I.}~\bibnamefont
  {Bischofberger}}, \bibinfo {author} {\bibfnamefont {R.}~\bibnamefont
  {Ramachandran}}, \ and\ \bibinfo {author} {\bibfnamefont {S.~R.}\
  \bibnamefont {Nagel}},\ }\bibfield  {title} {\enquote {\bibinfo {title}
  {Fingering versus stability in the limit of zero interfacial tension},}\
  }\href@noop {} {\bibfield  {journal} {\bibinfo  {journal} {Nat. Commun.}\
  }\textbf {\bibinfo {volume} {5}},\ \bibinfo {pages} {5265} (\bibinfo {year}
  {2014})}\BibitemShut {NoStop}%
\bibitem [{\citenamefont {Goyal}\ and\ \citenamefont
  {Meiburg}(2006)}]{Goyal06}%
  \BibitemOpen
  \bibfield  {author} {\bibinfo {author} {\bibfnamefont {N.}~\bibnamefont
  {Goyal}}\ and\ \bibinfo {author} {\bibfnamefont {E.}~\bibnamefont
  {Meiburg}},\ }\bibfield  {title} {\enquote {\bibinfo {title} {Miscible
  displacements in hele-shaw cells: two-dimensional base states and their
  linear stability},}\ }\href@noop {} {\bibfield  {journal} {\bibinfo
  {journal} {J. Fluid Mech.}\ }\textbf {\bibinfo {volume} {558}},\ \bibinfo
  {pages} {329--355} (\bibinfo {year} {2006})}\BibitemShut {NoStop}%
\bibitem [{\citenamefont {Oliveira}\ and\ \citenamefont
  {Meiburg}(2011)}]{Oliveira11}%
  \BibitemOpen
  \bibfield  {author} {\bibinfo {author} {\bibfnamefont {R.~M.}\ \bibnamefont
  {Oliveira}}\ and\ \bibinfo {author} {\bibfnamefont {E.}~\bibnamefont
  {Meiburg}},\ }\bibfield  {title} {\enquote {\bibinfo {title} {Miscible
  displacements in hele-shaw cells: three-dimensional navier--stokes
  simulations},}\ }\href@noop {} {\bibfield  {journal} {\bibinfo  {journal} {J.
  Fluid Mech.}\ }\textbf {\bibinfo {volume} {687}},\ \bibinfo {pages}
  {431--460} (\bibinfo {year} {2011})}\BibitemShut {NoStop}%
\bibitem [{\citenamefont {Tan}\ and\ \citenamefont {Homsy}(1987)}]{Tan87}%
  \BibitemOpen
  \bibfield  {author} {\bibinfo {author} {\bibfnamefont {C.~T.}\ \bibnamefont
  {Tan}}\ and\ \bibinfo {author} {\bibfnamefont {G.~M.}\ \bibnamefont
  {Homsy}},\ }\bibfield  {title} {\enquote {\bibinfo {title} {Stability of
  miscible displacements in porous media: Radial source flow},}\ }\href@noop {}
  {\bibfield  {journal} {\bibinfo  {journal} {Phys. Fluids}\ }\textbf {\bibinfo
  {volume} {30}},\ \bibinfo {pages} {1239--1245} (\bibinfo {year}
  {1987})}\BibitemShut {NoStop}%
\bibitem [{\citenamefont {Yortsos}(1987)}]{Yortsos87}%
  \BibitemOpen
  \bibfield  {author} {\bibinfo {author} {\bibfnamefont {Y.~C.}\ \bibnamefont
  {Yortsos}},\ }\bibfield  {title} {\enquote {\bibinfo {title} {Stability of
  displacement processes in porous media in radial flow geometries},}\
  }\href@noop {} {\bibfield  {journal} {\bibinfo  {journal} {Phys. Fluids}\
  }\textbf {\bibinfo {volume} {30}},\ \bibinfo {pages} {2928--2935} (\bibinfo
  {year} {1987})}\BibitemShut {NoStop}%
\bibitem [{\citenamefont {Bunton}\ \emph {et~al.}(2016)\citenamefont {Bunton},
  \citenamefont {Marin}, \citenamefont {Stewart}, \citenamefont {Meiburg},\
  and\ \citenamefont {De~Wit}}]{Bunton2016}%
  \BibitemOpen
  \bibfield  {author} {\bibinfo {author} {\bibfnamefont {P.}~\bibnamefont
  {Bunton}}, \bibinfo {author} {\bibfnamefont {D.}~\bibnamefont {Marin}},
  \bibinfo {author} {\bibfnamefont {S.}~\bibnamefont {Stewart}}, \bibinfo
  {author} {\bibfnamefont {E.}~\bibnamefont {Meiburg}}, \ and\ \bibinfo
  {author} {\bibfnamefont {A.}~\bibnamefont {De~Wit}},\ }\bibfield  {title}
  {\enquote {\bibinfo {title} {Schlieren imaging of viscous fingering in a
  horizontal hele-shaw cell},}\ }\href@noop {} {\bibfield  {journal} {\bibinfo
  {journal} {Exp. Fluids}\ }\textbf {\bibinfo {volume} {57}},\ \bibinfo {pages}
  {28} (\bibinfo {year} {2016})}\BibitemShut {NoStop}%
\bibitem [{\citenamefont {Yang}\ and\ \citenamefont {Yortsos}(1997)}]{Yang97}%
  \BibitemOpen
  \bibfield  {author} {\bibinfo {author} {\bibfnamefont {Z.}~\bibnamefont
  {Yang}}\ and\ \bibinfo {author} {\bibfnamefont {Y.~C.}\ \bibnamefont
  {Yortsos}},\ }\bibfield  {title} {\enquote {\bibinfo {title} {Asymptotic
  solutions of miscible displacements in geometries of large aspect ratio},}\
  }\href@noop {} {\bibfield  {journal} {\bibinfo  {journal} {Phys. Fluids}\
  }\textbf {\bibinfo {volume} {9}},\ \bibinfo {pages} {286--298} (\bibinfo
  {year} {1997})}\BibitemShut {NoStop}%
\bibitem [{\citenamefont {Rakotomalala}\ \emph {et~al.}(1997)\citenamefont
  {Rakotomalala}, \citenamefont {Salin},\ and\ \citenamefont
  {Watzky}}]{Rakotomalala97}%
  \BibitemOpen
  \bibfield  {author} {\bibinfo {author} {\bibfnamefont {N.}~\bibnamefont
  {Rakotomalala}}, \bibinfo {author} {\bibfnamefont {D.}~\bibnamefont {Salin}},
  \ and\ \bibinfo {author} {\bibfnamefont {P.}~\bibnamefont {Watzky}},\
  }\bibfield  {title} {\enquote {\bibinfo {title} {Miscible displacement
  between two parallel plates: Bgk lattice gas simulations},}\ }\href@noop {}
  {\bibfield  {journal} {\bibinfo  {journal} {J. Fluid Mech.}\ }\textbf
  {\bibinfo {volume} {338}},\ \bibinfo {pages} {277--297} (\bibinfo {year}
  {1997})}\BibitemShut {NoStop}%
\bibitem [{\citenamefont {Talon}\ \emph {et~al.}(2013)\citenamefont {Talon},
  \citenamefont {Goyal},\ and\ \citenamefont {Meiburg}}]{Talon13}%
  \BibitemOpen
  \bibfield  {author} {\bibinfo {author} {\bibfnamefont {L.}~\bibnamefont
  {Talon}}, \bibinfo {author} {\bibfnamefont {N.}~\bibnamefont {Goyal}}, \ and\
  \bibinfo {author} {\bibfnamefont {E.}~\bibnamefont {Meiburg}},\ }\bibfield
  {title} {\enquote {\bibinfo {title} {Variable density and viscosity, miscible
  displacements in horizontal hele-shaw cells. part 1. linear stability
  analysis},}\ }\href@noop {} {\bibfield  {journal} {\bibinfo  {journal} {J.
  Fluid Mech.}\ }\textbf {\bibinfo {volume} {721}},\ \bibinfo {pages}
  {268--294} (\bibinfo {year} {2013})}\BibitemShut {NoStop}%
\bibitem [{\citenamefont {Chuoke}\ \emph {et~al.}(1959)\citenamefont {Chuoke},
  \citenamefont {Van~Meurs},\ and\ \citenamefont {Van~der Poel}}]{Chuoke59}%
  \BibitemOpen
  \bibfield  {author} {\bibinfo {author} {\bibfnamefont {R.~L.}\ \bibnamefont
  {Chuoke}}, \bibinfo {author} {\bibfnamefont {P.}~\bibnamefont {Van~Meurs}}, \
  and\ \bibinfo {author} {\bibfnamefont {C.}~\bibnamefont {Van~der Poel}},\
  }\bibfield  {title} {\enquote {\bibinfo {title} {The instability of slow,
  immiscible, viscous liquid-liquid displacements in permeable media},}\
  }\href@noop {} {\bibfield  {journal} {\bibinfo  {journal} {Trans. Am. Inst.
  Min. Metall. Pet. Eng.}\ }\textbf {\bibinfo {volume} {216}},\ \bibinfo
  {pages} {188--194} (\bibinfo {year} {1959})}\BibitemShut {NoStop}%
\bibitem [{\citenamefont {Gardner}\ and\ \citenamefont
  {Ypma}(1984)}]{Gardner84}%
  \BibitemOpen
  \bibfield  {author} {\bibinfo {author} {\bibfnamefont {J.~W.}\ \bibnamefont
  {Gardner}}\ and\ \bibinfo {author} {\bibfnamefont {J.~G.~J.}\ \bibnamefont
  {Ypma}},\ }\bibfield  {title} {\enquote {\bibinfo {title} {An investigation
  of phase behavior-macroscopic bypassing interaction in co2 flooding},}\
  }\href@noop {} {\bibfield  {journal} {\bibinfo  {journal} {Soc. Pet. Eng.
  J.}\ }\textbf {\bibinfo {volume} {24}},\ \bibinfo {pages} {508--520}
  (\bibinfo {year} {1984})}\BibitemShut {NoStop}%
\bibitem [{\citenamefont {Liu}\ and\ \citenamefont {Nagel}(2010)}]{Liu2010}%
  \BibitemOpen
  \bibfield  {author} {\bibinfo {author} {\bibfnamefont {A.~J.}\ \bibnamefont
  {Liu}}\ and\ \bibinfo {author} {\bibfnamefont {S.~R.}\ \bibnamefont
  {Nagel}},\ }\bibfield  {title} {\enquote {\bibinfo {title} {The jamming
  transition and the marginally jammed solid},}\ }\href@noop {} {\bibfield
  {journal} {\bibinfo  {journal} {Annu. Rev. Condens. Matter Phys.}\ }\textbf
  {\bibinfo {volume} {1}},\ \bibinfo {pages} {347--369} (\bibinfo {year}
  {2010})}\BibitemShut {NoStop}%
\bibitem [{\citenamefont {Chen}(1987)}]{Chen1987}%
  \BibitemOpen
  \bibfield  {author} {\bibinfo {author} {\bibfnamefont {J.~D.}\ \bibnamefont
  {Chen}},\ }\bibfield  {title} {\enquote {\bibinfo {title} {Radial viscous
  fingering patterns in hele-shaw cells},}\ }\href@noop {} {\bibfield
  {journal} {\bibinfo  {journal} {Exp. Fluids}\ }\textbf {\bibinfo {volume}
  {5}},\ \bibinfo {pages} {363--371} (\bibinfo {year} {1987})}\BibitemShut
  {NoStop}%
\bibitem [{\citenamefont {Chen}(1989)}]{Chen89}%
  \BibitemOpen
  \bibfield  {author} {\bibinfo {author} {\bibfnamefont {J.~D.}\ \bibnamefont
  {Chen}},\ }\bibfield  {title} {\enquote {\bibinfo {title} {Growth of radial
  viscous fingers in a hele-shaw cell},}\ }\href@noop {} {\bibfield  {journal}
  {\bibinfo  {journal} {J. Fluid Mech.}\ }\textbf {\bibinfo {volume} {201}},\
  \bibinfo {pages} {223--242} (\bibinfo {year} {1989})}\BibitemShut {NoStop}%
\bibitem [{\citenamefont {Paterson}(1985)}]{Paterson85}%
  \BibitemOpen
  \bibfield  {author} {\bibinfo {author} {\bibfnamefont {L.}~\bibnamefont
  {Paterson}},\ }\bibfield  {title} {\enquote {\bibinfo {title} {Fingering with
  miscible fluids in a hele shaw cell},}\ }\href@noop {} {\bibfield  {journal}
  {\bibinfo  {journal} {Phys. Fluids}\ }\textbf {\bibinfo {volume} {28}},\
  \bibinfo {pages} {26--30} (\bibinfo {year} {1985})}\BibitemShut {NoStop}%
\bibitem [{\citenamefont {Aubertin}\ \emph {et~al.}(2009)\citenamefont
  {Aubertin}, \citenamefont {Gauthier}, \citenamefont {Martin}, \citenamefont
  {Salin},\ and\ \citenamefont {Talon}}]{Aubertin2009}%
  \BibitemOpen
  \bibfield  {author} {\bibinfo {author} {\bibfnamefont {A.}~\bibnamefont
  {Aubertin}}, \bibinfo {author} {\bibfnamefont {G.}~\bibnamefont {Gauthier}},
  \bibinfo {author} {\bibfnamefont {J.}~\bibnamefont {Martin}}, \bibinfo
  {author} {\bibfnamefont {D.}~\bibnamefont {Salin}}, \ and\ \bibinfo {author}
  {\bibfnamefont {L.}~\bibnamefont {Talon}},\ }\bibfield  {title} {\enquote
  {\bibinfo {title} {Miscible viscous fingering in microgravity},}\ }\href@noop
  {} {\bibfield  {journal} {\bibinfo  {journal} {Phys. Fluids}\ }\textbf
  {\bibinfo {volume} {21}},\ \bibinfo {pages} {054107} (\bibinfo {year}
  {2009})}\BibitemShut {NoStop}%
\bibitem [{\citenamefont {Ramachandran}(2017)}]{Ramachandran2017}%
  \BibitemOpen
  \bibfield  {author} {\bibinfo {author} {\bibfnamefont {R.}~\bibnamefont
  {Ramachandran}},\ }\bibfield  {title} {\enquote {\bibinfo {title} {Stability
  and onset of two-dimensional viscous fingering in immiscible fluids},}\
  }\href@noop {} {\bibfield  {journal} {\bibinfo  {journal} {arXiv preprint
  arXiv:1704.02674}\ } (\bibinfo {year} {2017})}\BibitemShut {NoStop}%
\bibitem [{\citenamefont {Paterson}(1981)}]{Paterson81}%
  \BibitemOpen
  \bibfield  {author} {\bibinfo {author} {\bibfnamefont {L.}~\bibnamefont
  {Paterson}},\ }\bibfield  {title} {\enquote {\bibinfo {title} {Radial
  fingering in a hele shaw cell},}\ }\href@noop {} {\bibfield  {journal}
  {\bibinfo  {journal} {J. Fluid Mech.}\ }\textbf {\bibinfo {volume} {113}},\
  \bibinfo {pages} {513--529} (\bibinfo {year} {1981})}\BibitemShut {NoStop}%
\bibitem [{\citenamefont {Nagel}\ and\ \citenamefont
  {Gallaire}(2013)}]{Nagel13}%
  \BibitemOpen
  \bibfield  {author} {\bibinfo {author} {\bibfnamefont {M.}~\bibnamefont
  {Nagel}}\ and\ \bibinfo {author} {\bibfnamefont {F.}~\bibnamefont
  {Gallaire}},\ }\bibfield  {title} {\enquote {\bibinfo {title} {A new
  prediction of wavelength selection in radial viscous fingering involving
  normal and tangential stresses},}\ }\href@noop {} {\bibfield  {journal}
  {\bibinfo  {journal} {Phys. Fluids}\ }\textbf {\bibinfo {volume} {25}},\
  \bibinfo {pages} {124107} (\bibinfo {year} {2013})}\BibitemShut {NoStop}%
\bibitem [{\citenamefont {Li}\ \emph {et~al.}(2009)\citenamefont {Li},
  \citenamefont {Lowengrub}, \citenamefont {Fontana},\ and\ \citenamefont
  {Palffy-Muhoray}}]{Li2009}%
  \BibitemOpen
  \bibfield  {author} {\bibinfo {author} {\bibfnamefont {S.}~\bibnamefont
  {Li}}, \bibinfo {author} {\bibfnamefont {J.~S.}\ \bibnamefont {Lowengrub}},
  \bibinfo {author} {\bibfnamefont {J.}~\bibnamefont {Fontana}}, \ and\
  \bibinfo {author} {\bibfnamefont {P.}~\bibnamefont {Palffy-Muhoray}},\
  }\bibfield  {title} {\enquote {\bibinfo {title} {Control of viscous fingering
  patterns in a radial hele-shaw cell},}\ }\href {\doibase
  10.1103/PhysRevLett.102.174501} {\bibfield  {journal} {\bibinfo  {journal}
  {Phys. Rev. Lett.}\ }\textbf {\bibinfo {volume} {102}},\ \bibinfo {pages}
  {174501} (\bibinfo {year} {2009})}\BibitemShut {NoStop}%
\bibitem [{\citenamefont {Al-Housseiny}\ \emph {et~al.}(2012)\citenamefont
  {Al-Housseiny}, \citenamefont {Tsai},\ and\ \citenamefont
  {Stone}}]{AlHousseiny12}%
  \BibitemOpen
  \bibfield  {author} {\bibinfo {author} {\bibfnamefont {T.~T.}\ \bibnamefont
  {Al-Housseiny}}, \bibinfo {author} {\bibfnamefont {P.~A.}\ \bibnamefont
  {Tsai}}, \ and\ \bibinfo {author} {\bibfnamefont {H.~A.}\ \bibnamefont
  {Stone}},\ }\bibfield  {title} {\enquote {\bibinfo {title} {Control of
  interfacial instabilities using flow geometry},}\ }\href@noop {} {\bibfield
  {journal} {\bibinfo  {journal} {Nat. Phys.}\ }\textbf {\bibinfo {volume}
  {8}},\ \bibinfo {pages} {747--750} (\bibinfo {year} {2012})}\BibitemShut
  {NoStop}%
\bibitem [{\citenamefont {Pihler-Puzovi{\'c}}\ \emph
  {et~al.}(2012)\citenamefont {Pihler-Puzovi{\'c}}, \citenamefont {Illien},
  \citenamefont {Heil},\ and\ \citenamefont {Juel}}]{Pihler2012}%
  \BibitemOpen
  \bibfield  {author} {\bibinfo {author} {\bibfnamefont {D.}~\bibnamefont
  {Pihler-Puzovi{\'c}}}, \bibinfo {author} {\bibfnamefont {P.}~\bibnamefont
  {Illien}}, \bibinfo {author} {\bibfnamefont {M.}~\bibnamefont {Heil}}, \ and\
  \bibinfo {author} {\bibfnamefont {A.}~\bibnamefont {Juel}},\ }\bibfield
  {title} {\enquote {\bibinfo {title} {Suppression of complex fingerlike
  patterns at the interface between air and a viscous fluid by elastic
  membranes},}\ }\href@noop {} {\bibfield  {journal} {\bibinfo  {journal}
  {Phys. Rev. Lett.}\ }\textbf {\bibinfo {volume} {108}},\ \bibinfo {pages}
  {074502} (\bibinfo {year} {2012})}\BibitemShut {NoStop}%
\bibitem [{\citenamefont {Zheng}\ \emph {et~al.}(2015)\citenamefont {Zheng},
  \citenamefont {Kim},\ and\ \citenamefont {Stone}}]{Zheng2015}%
  \BibitemOpen
  \bibfield  {author} {\bibinfo {author} {\bibfnamefont {Z.}~\bibnamefont
  {Zheng}}, \bibinfo {author} {\bibfnamefont {H.}~\bibnamefont {Kim}}, \ and\
  \bibinfo {author} {\bibfnamefont {H.~A.}\ \bibnamefont {Stone}},\ }\bibfield
  {title} {\enquote {\bibinfo {title} {Controlling viscous fingering using
  time-dependent strategies},}\ }\href {\doibase
  10.1103/PhysRevLett.115.174501} {\bibfield  {journal} {\bibinfo  {journal}
  {Phys. Rev. Lett.}\ }\textbf {\bibinfo {volume} {115}},\ \bibinfo {pages}
  {174501} (\bibinfo {year} {2015})}\BibitemShut {NoStop}%
\bibitem [{\citenamefont {D'Errico}\ \emph {et~al.}(2004)\citenamefont
  {D'Errico}, \citenamefont {Ortona}, \citenamefont {Capuano},\ and\
  \citenamefont {Vitagliano}}]{dErrico2004}%
  \BibitemOpen
  \bibfield  {author} {\bibinfo {author} {\bibfnamefont {G.}~\bibnamefont
  {D'Errico}}, \bibinfo {author} {\bibfnamefont {O.}~\bibnamefont {Ortona}},
  \bibinfo {author} {\bibfnamefont {F.}~\bibnamefont {Capuano}}, \ and\
  \bibinfo {author} {\bibfnamefont {V.}~\bibnamefont {Vitagliano}},\ }\bibfield
   {title} {\enquote {\bibinfo {title} {Diffusion coefficients for the binary
  system glycerol+ water at 25 c. a velocity correlation study},}\ }\href@noop
  {} {\bibfield  {journal} {\bibinfo  {journal} {J. Chem. Eng. Data}\ }\textbf
  {\bibinfo {volume} {49}},\ \bibinfo {pages} {1665--1670} (\bibinfo {year}
  {2004})}\BibitemShut {NoStop}%
\bibitem [{\citenamefont {Petitjeans}\ and\ \citenamefont
  {Maxworthy}(1996)}]{Petitjeans1996}%
  \BibitemOpen
  \bibfield  {author} {\bibinfo {author} {\bibfnamefont {P.}~\bibnamefont
  {Petitjeans}}\ and\ \bibinfo {author} {\bibfnamefont {T.}~\bibnamefont
  {Maxworthy}},\ }\bibfield  {title} {\enquote {\bibinfo {title} {Miscible
  displacements in capillary tubes. part 1. experiments},}\ }\href@noop {}
  {\bibfield  {journal} {\bibinfo  {journal} {J. Fluid Mech.}\ }\textbf
  {\bibinfo {volume} {326}},\ \bibinfo {pages} {37--56} (\bibinfo {year}
  {1996})}\BibitemShut {NoStop}%
\bibitem [{\citenamefont {Cheng}(2008)}]{Cheng2008}%
  \BibitemOpen
  \bibfield  {author} {\bibinfo {author} {\bibfnamefont {N.-S.}\ \bibnamefont
  {Cheng}},\ }\bibfield  {title} {\enquote {\bibinfo {title} {Formula for the
  viscosity of a glycerol- water mixture},}\ }\href@noop {} {\bibfield
  {journal} {\bibinfo  {journal} {Ind. Eng. Chem. Res.}\ }\textbf {\bibinfo
  {volume} {47}},\ \bibinfo {pages} {3285--3288} (\bibinfo {year}
  {2008})}\BibitemShut {NoStop}%
\end{thebibliography}%

\end{document}